\documentclass[pra,aps,amsmath,amssymb,amsfonts,twocolumn,nofootinbib,longbibliography,superscriptaddress]{revtex4-2}
\usepackage{amssymb}
\usepackage{bm,mathrsfs}
\usepackage{graphicx}
\usepackage{epsfig}
\usepackage{amsmath,bbm}
\usepackage{amsfonts,amssymb}
\usepackage{times}
\usepackage{verbatim}
\usepackage[sort&compress]{natbib}
\usepackage{amsmath}
\usepackage{bm}
\usepackage{float}
\usepackage{booktabs}
\usepackage{colortbl}
\usepackage{graphicx}
\usepackage{amsmath}
\usepackage{amssymb}
\usepackage{braket}
\usepackage{array}
\usepackage{makecell}
\usepackage{multirow}
\newcommand{\gmp}{\gamma_{p}}
\newcommand{\omp}{\Omega_{p}}
\newcommand{\kpp}{\kappa}

\usepackage{xcolor}
\definecolor{myblue}{rgb}{0.00,0.20,0.60}
\allowdisplaybreaks[4]
\usepackage[colorlinks,breaklinks,linkcolor=blue,anchorcolor=blue,citecolor=blue,urlcolor=blue]{hyperref}

\begin{document}     
\title{Floquet-engineered dissipative stabilizer pumping in neutral atom arrays}
\author{F. Q. Guo}
\affiliation{Center for Quantum Science and School of Physics, Northeast Normal University, Changchun 130024, China}

\author{Shi-Lei Su}
\email{slsu@zzu.edu.cn}
\affiliation{School of Physics, Key Laboratory of Materials Physics of Ministry of Education, and International Laboratory for Quantum Functional Materials of Henan, Zhengzhou University, Zhengzhou 450001, China}
\affiliation{Institute of Quantum Materials and Physics, Henan Academy of Science, Henan 450046, China}

\author{Weibin Li}
\email{weibin.li@nottingham.ac.uk}
\affiliation{School of Physics and Astronomy, and Centre for the Mathematics and Theoretical Physics of Quantum Nonequilibrium Systems, The University of Nottingham, Nottingham NG7 2RD, United Kingdom}

\author{X. Q. Shao}
\email{xqshao@nenu.edu.cn}
\affiliation{Center for Quantum Science and School of Physics, Northeast Normal University, Changchun 130024, China}
\affiliation{Institute of Quantum Science and Technology, Yanbian University, Yanji 133002, China}

\begin{abstract}
Autonomous stabilization of many-body quantum states requires a mechanism that continuously removes stabilizer violations without relying on measurement and feedback. Here we develop a Floquet-Lindblad framework for scalable stabilizer pumping in neutral Rydberg atom arrays. Periodically modulated Rydberg excitation selectively couples the negative-eigenvalue sector of a prescribed stabilizer to auxiliary Rydberg states, while engineered dissipation irreversibly transfers the population into the corresponding positive-eigenvalue sector and leaves the target stabilizer manifold dark. This establishes a modular pumping primitive that can be composed to autonomously stabilize general many-body stabilizer manifolds, rather than merely prepare a specific entangled state. The combination of Floquet-selective excitation and fast engineered dissipation accelerates convergence, while transient single-Rydberg excitation and suppressed real multi-Rydberg occupation reduce sensitivity to Rydberg-state loss, blackbody-radiation-induced leakage, and spatial fluctuations. We demonstrate the framework through Bell and multiqubit graph states, including parallel stabilizer pumping that reduces the convergence time for the simulated multiqubit systems. The same dissipative attraction also enables purification of imperfect resource states. These results establish Floquet-engineered stabilizer pumping as a scalable route toward autonomous quantum-state stabilization and protected entanglement generation in neutral-atom platforms.
\end{abstract}

\maketitle

\section{Introduction}

Reservoir engineering of qubit–environment coupling provides a powerful route to autonomously steering quantum systems toward target many-body states and stabilizing them as nonequilibrium steady states~\cite{Cho2011Optical,PhysRevLett.106.090502,Rao2013Dark,Bentley2014Detection,Morigi2015Dissipative,Reiter2016Scalable,Qin2018Exponentially,Harrington2022Quantum,Shah2024Stabilizing,Vivas2024Frequency,zhu2025dissipation,PhysRevLett.134.050603}. Among the most important targets are stabilizer states, which provide central resources for quantum information processing and offer structured many-body manifolds for exploring engineered driven and open-system dynamics~\cite{Raussendorf2001A,Kraus2008Preparation,Barreiro2011An,Muller2011Simulating}. While autonomous entanglement stabilization has been experimentally demonstrated in few-qubit trapped-ion~\cite{Barreiro2011An,Lin2013Dissipative,Cole2022Resource,Malinowski2022Generation} and superconducting platforms~\cite{Shankar2013Autonomously,Liu2016Comparing,Kimchi2016Stabilizing,chen2024hardware}, extending such dissipative control to many-body stabilizer manifolds remains challenging because the required pumping processes must be selective, composable, and scalable. Neutral atoms offer a particularly promising platform for scalable implementations~\cite{Isenhower2010Demonstration,Levine2019Parallel,Evered2023High,Ma2023High,Daniel2016An,Bernien2017Probing,Sylvain2019Observation,Browaeys2020Many,Lienhard2020Realization,Bluvstein2022A,Su2023Dipolar,Zeybek2023Quantum,bai2026multipartite}, owing to deterministic atom-by-atom assembly in optical tweezers~\cite{Saffman2010Quantum,Saffman2016Quantum,Morgado2021Quantum,Wu2021A,shao2024rydberg}, strong long-range interactions, and precise control over electronic and motional degrees of freedom~\cite{PhysRevLett.123.213603,Ebadi2021Quantum,chen2023continuous,PhysRevX.14.011025,chen2025spectroscopy,w1cp-l5vq,5qhh-322q}. 
However, incoherent processes—such as finite lifetimes of electronic states and blackbody radiation (BBR)~\cite{Saffman2010Quantum}—as well as imperfections in array structures~\cite{soto-garciaInfiniteRandomnessCriticality2026}, inevitably perturb the coherent and dissipative dynamics required for stabilizing many-body quantum states. These limitations have motivated extensive studies of open-system dynamics in Rydberg platforms~\cite{begoc2023controlled,zhouDiagnosingQuantumManybody2025,chen2025,jptr-pm37}, including dissipative entanglement schemes based predominantly on antiblockade-assisted excitation~\cite{Carr2013Preparation,Shao2014Stationary,Su2015Simplified,Roghani_2018,Wintermantel202Unitary,Varghese2023Maximally}. In such schemes, dissipative transitions are typically engineered for specific target states and rely on resonance conditions set by the interatomic interaction strength. Such approaches require precise control of the interatomic spacing $R$, rendering the protocols highly sensitive to spatial disorder. A key challenge is therefore to develop a modular dissipative mechanism that can selectively pump stabilizer-violating sectors into a prescribed stabilizer manifold, while remaining scalable, robust against realistic imperfections, and insensitive to stringent interaction-matching conditions.

\begin{figure}
	\centering
	\includegraphics[width=1\linewidth]{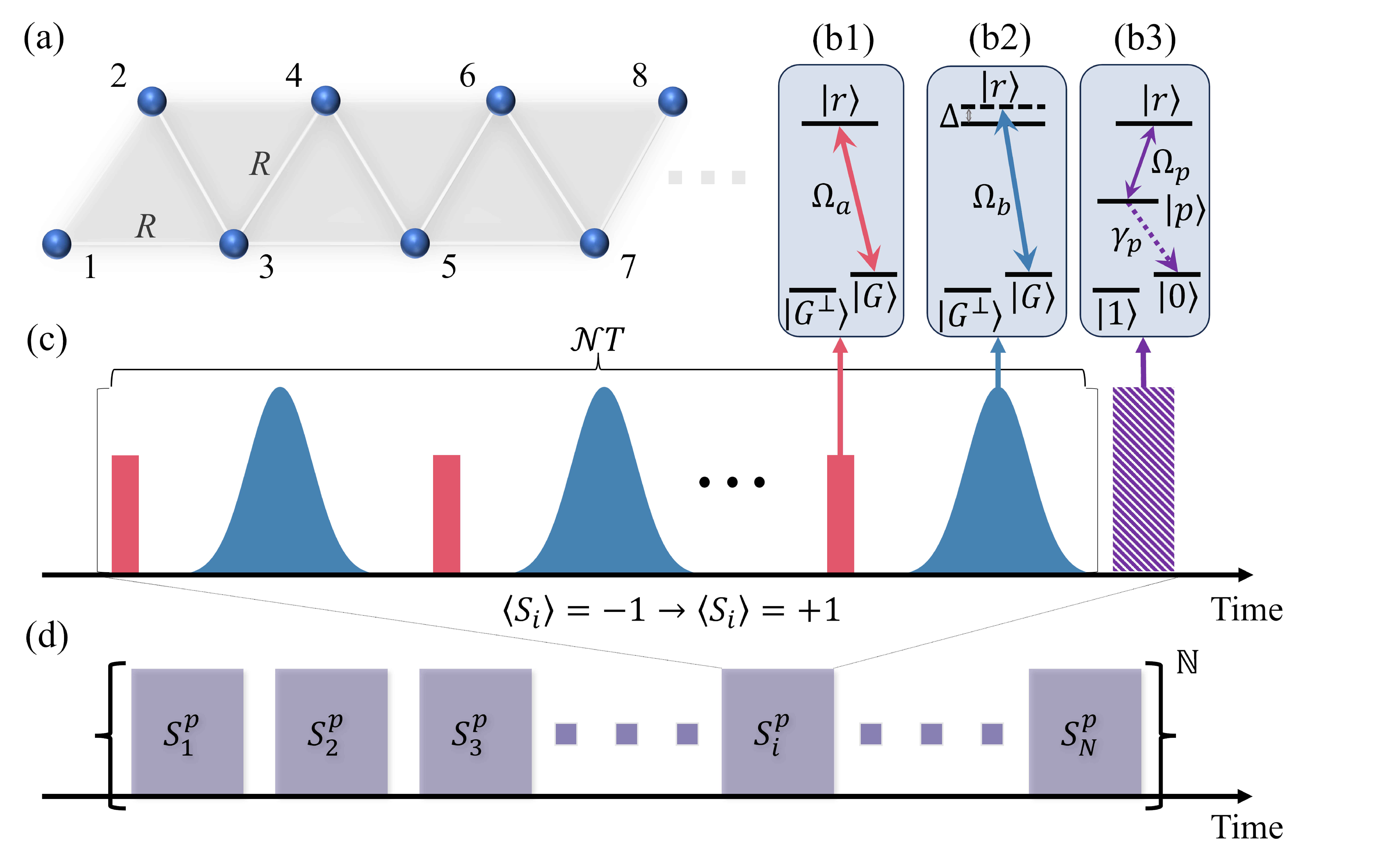}
	\caption{
(a) Zig-zag array of atoms (blue dots), where each set of three neighboring atoms forms an equilateral triangle with nearest-neighbor spacing $R$.
(b) Level diagrams illustrating the resonant \(\pi/\mathcal{N}\) pulse (b1), the Gaussian pulse (b2), and the engineered dissipative reset via the intermediate state (b3).  The ground state $|G\rangle$ is defined as the eigenstate of either the Pauli-$Z$ or Pauli-$X$ operator within the ground-state subspace, corresponding to an eigenvalue of $-1$. The orthogonal state $|G^{\perp}\rangle$ remains decoupled from the dynamics.
(c) Stabilizer pumping process \(S_i^p\) for generator \(S_i\). Each cycle comprises a Floquet-selective coherent-excitation stage (duration \(\mathcal{N}T\)) followed by engineered dissipative reset, which together induce unidirectional population transfer from the \( S_i =-1\) eigenspace to the \( S_i = +1\) eigenspace.
(d) The sequence of composable stabilizer-pumping operations.}
\label{fig1}
\end{figure}

In this work, we propose a Floquet-Lindblad scheme that addresses these limitations  by leveraging periodic driving to engineer a robust dissipative landscape for stabilizer pumping. Our approach realizes stabilizer pumping toward many-body stabilizer states in a two-dimensional zig-zag atom array [Fig.~\ref{fig1}(a)], maintaining high fidelity even in the presence of spatial disorder. The dynamics of the atomic states [Fig.~\ref{fig1}(b)] alternate between two independent coherent and dissipative stages: a noninstantaneous unitary kick drives selective transitions between atomic ground states (qubits), suppressing position-dependent noise and Doppler dephasing, while engineered dissipation is exploited to drive population toward the target stabilizer sector and accelerate the pumping process. This framework also naturally accommodates extensions such as recycling BBR-induced leakage back into the computational subspace. Unlike approaches relying on purely unitary state preparation or real-time measurement and feedback~\cite{Bluvstein2022A, Liu2016Comparing}, our scheme autonomously drives the system toward many-body stabilizer manifolds that form dark steady states. Such independent control over coherent and dissipative segments provides both robustness and experimental versatility. The resulting dynamics are insensitive to the initial state and naturally support the composition of multiple stabilizer-pumping processes. This capability underpins scalable preparation, purification, and protection of stabilizer states, opening new avenues for measurement-based quantum computation (MBQC).

\section{Model and scheme}

The system consists of $N$ atoms arranged in a zig-zag geometry in the $x$-$y$ plane with uniform nearest-neighbor spacing $R$ [Fig.~\ref{fig1}(a)], individually addressed by laser fields~\cite{Manuel2016Atom, Daniel2016An, Barredo2018Synthetic, Graham2022Multi, Graham2023Multiscale, Evered2023High, Bluvstein2024Logical, Radnaev2025Universal}. To implement the coherent stage of stabilizer pumping, the atomic ground state $|G\rangle$ is coupled to a Rydberg state $|r\rangle$ via two alternating driving operations: a short resonant $\pi/\mathcal{N}$ pulse [Fig.~\ref{fig1}(b1)] followed by a blue-detuned $\pi$ pulse [Fig.~\ref{fig1}(b2)]. The Hamiltonian governing these processes is
\begin{equation}
     H_{\chi} = \sum_{i=1}^N\frac{\Omega_{\chi}}{2}e^{-i\delta_{{\chi,b}}\Delta t}|r_{i}\rangle\langle G_{i}| + \mathrm{H.c.} + \sum_{i< j}{U^{(ij)}_{rr}}|r_{i}r_j\rangle\langle r_{i}r_j|,
\end{equation}
where $\delta_{\chi,b}$ ($\chi = a, b$) denotes the Kronecker delta. Upon excitation to the Rydberg state, atoms at sites $i$ and $j$ interact via the van der Waals (vdW) potential $U^{(ij)}_{rr}=-C_6/R_{ij}^6$, which is determined by the interatomic distance $R_{ij}$ and the dispersion coefficient $C_6$. The Rabi frequency \(\Omega_a\) follows a square-pulse profile, whereas \(\Omega_b\) employs a Gaussian envelope to suppress unwanted off-resonant transitions; see Appendix~\ref{app:gaussian} for details. The Gaussian profile is given by \(\Omega_b(t) = \Omega \exp\left[-(t - 2t_f)^2 / (\sqrt{2} \alpha t_f)^2\right]\), with a peak amplitude \(\Omega\), effective pulse width \(\alpha t_f\), and center \(2t_f\). Here, \(t_f = \sqrt{\pi/2} / [\alpha \Omega \mathrm{erf}(\sqrt{2}/\alpha)]\), and the total duration is \(t_b=4t_f\).

For a target state \(|\psi\rangle\), a Pauli-string operator \(S_i\) satisfying \(S_i|\psi\rangle=|\psi\rangle\) defines a stabilizer constraint on \(|\psi\rangle\)~\cite{Toth2005Entanglement}. Building on this stabilizer structure, we employ a stabilizer pumping scheme to drive the system toward the target stabilizer manifold~\cite{Barreiro2011An, Cho2011Optical, Muller2011Simulating, Reiter2016Scalable}. This is achieved by alternating the coherent and dissipative stages for \(\mathcal{N}\) cycles. Specifically, eigenstates of the stabilizer generator \(S_i\) with eigenvalue \(-1\) (stabilizer-violating states) are selectively excited to the Rydberg state and subsequently decay back to the qubit subspace via an engineered dissipation channel [Fig.~\ref{fig1}(b3)]. This process, denoted as \(S_i^p\), induces a unidirectional population flow from the negative eigenspace ($S_i = -1$) to the positive eigenspace ($S_i = +1$), as illustrated in Fig.~\ref{fig1}(c). Each operation results in partial population transfer into the positive eigenspace. Through repeated cycles, the negative eigenspace components are gradually depleted, and the system asymptotically converges to a steady state residing in the joint positive eigenspace of all stabilizer generators [Fig.~\ref{fig1}(d)].

\begin{figure}
	\centering
   	\includegraphics[width=1\linewidth]{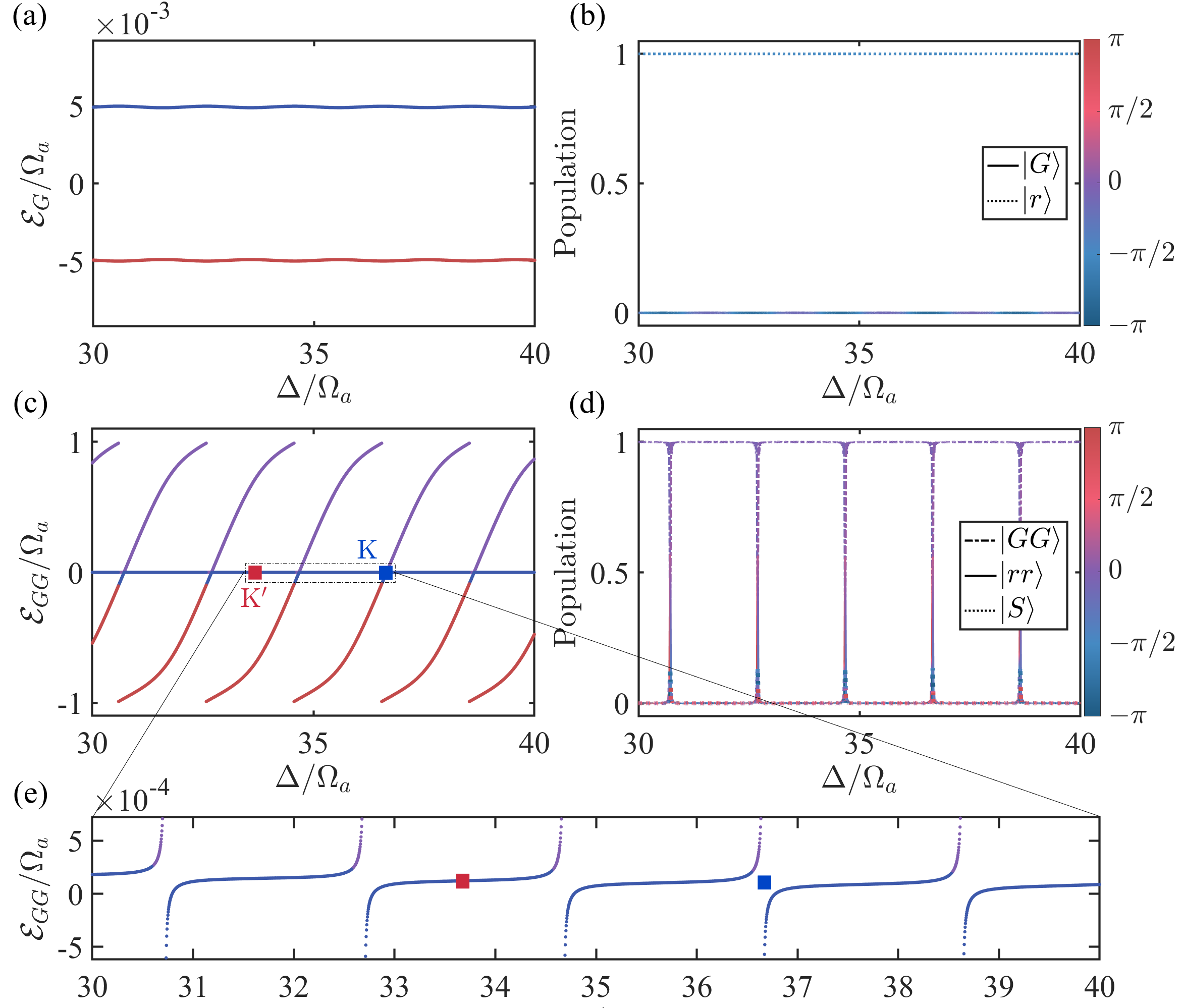}
	\caption{(a) Quasienergies of the Floquet Hamiltonian $H_F$, which reduces to an effective single-atom model within the driven subspace where the interaction term vanishes. (b) State populations for a single atom initialized in $|G\rangle$, plotted as a function of detuning $\Delta$. The detuning is scaled to the interaction strength $U_{rr}$ used in the two-atom analysis to enable a direct comparison. 
(c, d) Corresponding quasienergy $\mathcal{E}_{GG}$ and populations for a two-atom system initialized in $|GG\rangle$, where the symmetric state is defined as $|S\rangle = (|Gr\rangle + |rG\rangle)/\sqrt{2}$. The system exhibits a series of avoided crossings (e.g., point ${\rm K}$ at $\Delta/\Omega_a = 36.655$) and regular points (e.g., ${\rm K'}$ at $\Delta/\Omega_a = 33.68$). 
(e) Enlarged view of the spectrum around points ${\rm K}$ and ${\rm K'}$. In (b) and (d), the curve colors represent the phase angles of the probability amplitudes for each quantum state. Parameters are set to $\Omega/\Omega_{a} = 2.6619$, $t_a = \pi/(\mathcal{N}\Omega_a)$, $t_b = \mathcal{N}t_a$, $\alpha=0.6$, and $\mathcal{N} = 100$.}\label{fig2}
\end{figure}

\subsection{Floquet-selective excitation}

To elucidate the physical mechanism, we first analyze a two-atom system. The stroboscopic dynamics of the coherent Floquet driving are described by the time-independent Floquet Hamiltonian $H_{F} = ({i}/{T})\log[\mathcal{T}\exp(-i\int_0^{t_b}H_{b}dt)\exp({-iH_at_a)}]$, where $\mathcal{T}$ represents time ordering, and $T=t_a+t_b$ is the driving period with $t_a=\pi/(\mathcal{N}\Omega_a)$~\cite{bukov2015universal,IVANOV202117}. 
For the Gaussian pulse $\Omega_b$, the duration is fixed at $t_b=\mathcal{N}t_a$ ($\mathcal{N}\gg1$) to impart a strong non-impulsive kick. For simplicity, its time-averaged Rabi frequency during the interaction is set to equal $\Omega_a$. 
In the regime of large detuning and strong interaction ($\Delta= U_{rr}^{(12)} \gg \Omega_{a,b}$), and in the large $\mathcal{N}$ limit, the dynamics are confined to the Zeno subspace, governed by an effective Hamiltonian (see Appendix~\ref{app:zeno})~\cite{Facchi2004Unification, Facchi2008Quantum,Shao2011Quantum}
\begin{equation}\label{eff}
H_{\rm eff}\approx
\sum_{i\neq j}\frac{\Omega_{\rm eff}}{2}|r_i\rangle\langle G_i|\otimes|G_j^{\perp}\rangle\langle G_j^{\perp}|+\mathrm{H.c.},
\end{equation}
where the effective Rabi frequency is $\Omega_{\rm eff}=\Omega_a t_a/T$. Here, $|G^{\perp}\rangle$ denotes a qubit state orthogonal to \(|G\rangle\) that decouples from the dynamics (see Appendix~\ref{app:zeno}).
Equation~(\ref{eff}) reveals that the dynamics are kinetically constrained by the neighbor's state.  The system exhibits single-atom Rabi oscillations when initialized in $|G_i G_j^\perp\rangle$ (one atom in $|G\rangle$, the other in $|G^\perp\rangle$), but remains frozen for the state $|GG\rangle$, thereby providing the state selectivity required for stabilizer pumping.

This picture is verified by numerically simulating the Floquet Hamiltonian, with results presented in Fig.~\ref{fig2}. Figure~\ref{fig2}(a) shows that for the effective single-atom model (neglecting the decoupled state $|G^{\perp}\rangle$), the quasienergy spectrum exhibits a constant gap. This signifies a steady driving strength, resulting in coherent Rabi oscillations between $|G\rangle$ and $|r\rangle$ at frequency $|\Omega_{\rm eff}|$ (see Appendix~\ref{app:gaussian}). The physical origin is that the Floquet eigenstates $|\varepsilon_G^{\pm}\rangle = ({|G\rangle \pm |r\rangle})/\sqrt{2}$ are quasi-stationary, such that their projections remain conserved during the stroboscopic evolution. Figure~\ref{fig2}(b) confirms that a perfect exchange between $|G\rangle$ and $|r\rangle$ is achieved at the end of the pulse ($t_{\rm end}=\pi/\Omega_{\rm eff}$).

With both atoms initially in $|G\rangle$ and simultaneously driven, the dynamics are primarily governed by $H_b$, which possesses an intrinsic period $2\pi/\Delta$ and a quasienergy bandwidth $\Delta$. However, imposing an external period $T$ via pulse engineering creates a mismatch with the intrinsic periodicity. Consequently, the original quasienergy spectrum ``folds'' into the new Floquet Brillouin zone defined by $T$, causing an aliasing effect [Fig.~\ref{fig2}(c)]. Hybridization between the two-atom ground and Rydberg-dressed states manifests as avoided crossings in \(\mathcal{E}_{GG}\). These crossings occur near \(\Delta \cdot T = (2m+1)\pi\) (\(m \in \mathbb{Z}\)) and repeat with a period determined by the Floquet frequency \(2\pi/T\)~(see Appendix~\ref{app:avoided}).
Away from these points, the dynamics align well with Eq.~(\ref{eff}) [see Fig.~\ref{fig2}(d)], confirming that the Floquet dynamics selectively excite ground-state atoms conditioned on the partner occupying a decoupled state.

\subsection{Engineered dissipative reset}

Following the coherent excitation stage, we now investigate the engineered dissipative reset that completes each stabilizer-pumping cycle. A continuous laser field couples the Rydberg state to an intermediate level $|p\rangle$ with decay rate $\gamma_p$ [Fig.~\ref{fig1}(b3)].
The Liouvillian of this three-level system possesses nine eigenmodes with eigenvalues $\lambda_0,\lambda_1,\ldots,\lambda_8$.
The zero-eigenvalue mode ($\lambda_0=0$) corresponds to the steady state $\rho_{\mathrm{ss}}=|0\rangle\langle0|$. Details of the Liouvillian spectrum can be found in Appendix~\ref{app:reset}.

\begin{figure}
	\centering
	\includegraphics[width=0.96\linewidth]{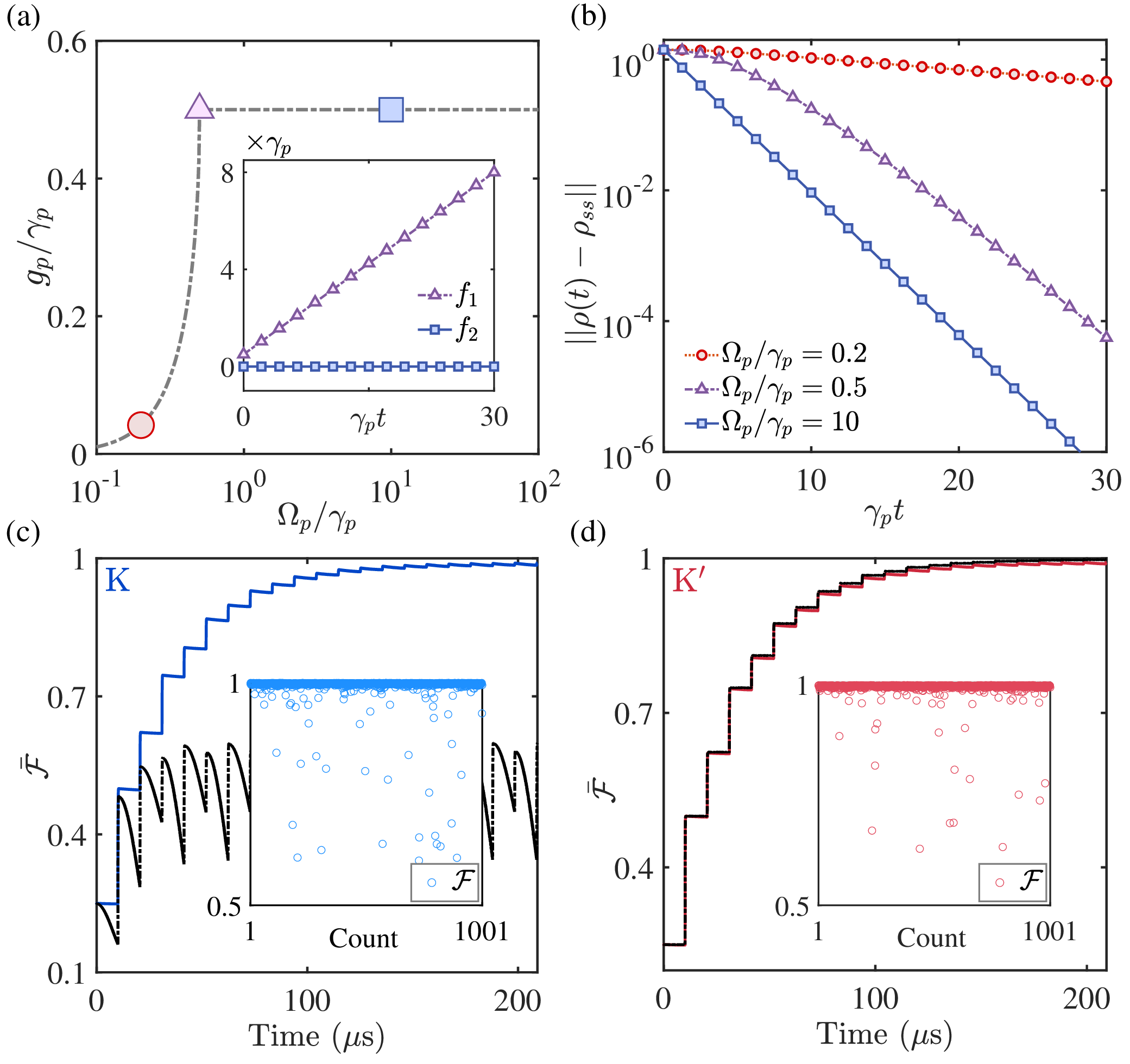}
	\caption{(a) Real part of the Liouvillian spectral gap $g_p = {\rm Re}[\gamma_p - \sqrt{\gamma_p^2 - 4\Omega_p^2}]/2$. Inset: time derivatives of non-steady state populations with the common factor $\exp(-\gamma_p t/2)$ removed. At the Liouvillian exceptional point (LEP), the derivative $f_1>0$ indicates a decay slower than that achieved in the strong-driving regime ($f_2=0$). 
(b) Relaxation of the Hilbert-Schmidt distance
$\|\rho(t)-\rho_{\mathrm{ss}}\|$ during the engineered dissipative reset shown in Fig.~\ref{fig1}(b3). The curves correspond to the parameter regimes indicated by the symbols in panel (a).
(c, d) Bell-state stabilizer-pumping fidelity at the avoided-crossing point ${\rm K}$ and regular point ${\rm K'}$. Dashed lines represent zero-temperature simulations, and solid lines show results at $5~\mu{\rm K}$ which are averaged over $1001$ random spatial realizations (distributions in insets). 
Points ${\rm K}$ and ${\rm K'}$ correspond to interatomic distances $R = 5.3382~\mu\text{m}$ and $5.4141~\mu\text{m}$, respectively. Other parameters: $\Omega_a = 2\pi\times 5~\mathrm{MHz}$, $\mathcal{N}=100$, $\alpha = 0.6$, $t_b = \mathcal{N}t_a = 0.1~\mu\text{s}$, $\mathbb{N} = 10$, $\Omega \approx 2\pi\times 13.31~{\rm MHz}$, $\gamma_p = 2\pi\times 6.07~\mathrm{MHz}$, decay duration $t_d = 0.362~\mu{\rm s}$.}\label{bell}
\end{figure}

The principle of accelerated dissipation parallels the quantum Mpemba effect~\cite{Carollo2021Exponentially,Chatterjee2023Quantum,Joshi2024Observing,Nava2024Mpemba,Zhang2025Observation}, as the initial state (Rydberg state $|r\rangle$, populated via the preceding coherent excitation) overlaps exclusively with the rapidly decaying eigenmodes ($\lambda_5$, $\lambda_6$, and $\lambda_8$).
The decay rate is thus governed by the smallest  spectral gap, $\text{min}\{|{\rm Re}(\lambda_{5})|,|{\rm Re}(\lambda_{6})|,|{\rm Re}(\lambda_{8})|\}$, which, for this case, admits the analytical form $g_p={\rm Re}[\gamma_p-\sqrt{\gamma_p^2-4\Omega_p^2}]/2$.
As shown in Fig.~\ref{bell}(a), $g_p$ is tunable via $\Omega_p$. In the conventional adiabatic regime $\Omega_p \ll \gamma_p$~\cite{begoc2023controlled,chen2025,jptr-pm37} (circle), the decay rate scales as $g_p \approx \Omega_p^2/\gamma_p$. Although faster than Rydberg spontaneous emission, this remains significantly slower than $\gamma_p$. Near the Liouvillian exceptional point (LEP) at $\Omega_p = \gamma_p/2$ (triangle)~\cite{Zhang2025Observation,Zhou2023Accelerating}, one might anticipate purely exponential convergence at a rate of $\gamma_p/2$. However, the ideal exponential decay is not realized near the LEP; it emerges only in the strong-driving limit $\Omega_p \gg \gamma_p$ (square).

Rigorous analysis in the biorthogonal basis uncovers a time-dependent prefactor at the LEP that grows over time, slowing the relaxation~(see Appendix~\ref{app:reset}). In the inset of Fig.~\ref{bell}(a), the time derivatives of the non-steady state populations are shown, with \(\exp(-\gamma_p t/2)\) factored out. At the LEP, the residual term \(f_1 = {\gamma_p}(1 + \gamma_p t/2)/2\) is non-negligible and increases with time. Conversely, in the strong-driving regime, this residual term essentially vanishes ($f_2 \approx 0$), facilitating faster relaxation.
As shown in Fig.~\ref{bell}(b), this yields exponential convergence in the limit $\Omega_p \gg \gamma_p$,
\begin{equation}
     ||\rho(t)-\rho_{ss}||\approx \sqrt{2}e^{-\gamma_pt/2},
\end{equation}
where $||\sigma||=\sqrt{{\rm Tr[\sigma\sigma^{\dagger}]}}$ is the Hilbert-Schmidt norm, $\rho(t)$ is the system density matrix, and $\rho_{ss} = |0\rangle \langle 0|$.
The factor $\sqrt{2}$ arises from the orthogonality between the initial and target states.
Since $\gamma_p$ is in the MHz range, relaxation occurs on a submicrosecond timescale—significantly shorter than the Rydberg lifetime.
This rapid relaxation also helps mitigate quantum information loss induced by BBR at room temperature (300~K). As demonstrated by the numerical simulations in Appendix~\ref{app:bbr}, this method effectively recycles BBR-induced leakage back into the computational subspace. In particular, accounting for the four dominant BBR-induced Rydberg states, sequentially recycling these populations adds less than $1.5~\mu\mathrm{s}$ to each dissipative-reset stage.

\section{Composition of stabilizer pumps}

We first demonstrate the composition of two stabilizer-pumping operations using the bipartite Bell state as the minimal example, $|\Phi^+\rangle = (|0_1 0_2\rangle + |1_1 1_2\rangle)/\sqrt{2}$. As a common $+1$ eigenstate of stabilizers $Z_1 Z_2$ and $X_1 X_2$, $|\Phi^+\rangle$ can be reached from arbitrary initial states by sequentially pumping the two stabilizers [Fig.~\ref{fig1}(d)]. Specifically, the $-1$ eigenspaces of the two stabilizers are sequentially coupled to the Rydberg state $|r\rangle$: first targeting the $Z_{1}Z_{2}$ stabilizer by substituting $|G_i\rangle$ with $|1_i\rangle$ in Eq.~(\ref{eff}), and subsequently targeting $X_{1}X_{2}$ by substituting $|G_i\rangle$ with $|-_i\rangle$. These addressed subspaces correspond to the $-1$ eigenspaces of the respective stabilizer generators.

Our proposal is compatible in principle with either alkali or alkaline-earth atoms.  To be concrete, we illustrate the scheme with a pair of $^{87}\mathrm{Rb}$ atoms. The qubit is encoded in hyperfine states $|0\rangle \equiv |5S_{1/2}, F=2, m_F=2\rangle$ and $|1\rangle \equiv |5S_{1/2}, F=1, m_F=0\rangle$.
The Rydberg state $|r\rangle \equiv |80S_{1/2}\rangle$ provides strong and isotropic van der Waals interactions, characterized by $C_6/(2\pi) = -4241.25~\mathrm{GHz}~(\mu\mathrm{m})^6$~\cite{SIBALIC2017319}. Dissipation is engineered via the intermediate state  $|p\rangle$ ($5P_{3/2}$)~(see Appendix~\ref{app:experiment}). To show independence from the initial state, the system is initialized as the maximally mixed state in the computational subspace, denoted as $\rho(0) = \sum_{i,j=0,1}|ij\rangle\langle ij|/4$. The stabilizer-pumping dynamics toward the Bell state are then examined in the ideal zero-temperature regime, neglecting the thermal fluctuations. Two typical situations are considered: one where the parameters correspond to the avoided-crossing point ${\rm K}$ and the other to a regular point ${\rm K'}$ [Fig.~\ref{fig2}(c)]. Figures~\ref{bell}(c) and \ref{bell}(d) show the time evolution of the Bell-state fidelity under the Floquet-Lindblad dynamics $\mathcal{F}=\langle \Phi^+|\rho(t)|\Phi^+\rangle$ (black dashed lines),  highlighting the distinct dynamical behaviors in the two regimes.
At the narrow avoided-crossing point ${\rm K}$, the state $|GG\rangle$ strongly hybridizes with nearby excited states during the pumping process, breaking the desired Floquet selectivity and thereby degrading the fidelity.
At ${\rm K'}$, by contrast, the fidelity reaches as high as $99.25\%$ after $\mathbb{N}=10$ pumping cycles.
In this regime, the state $|GG\rangle$ is separated from the other states by a sizable quasienergy gap, which effectively suppresses undesired hybridization and preserves high fidelity.

Notably, the protocol exhibits remarkable resilience to the thermal fluctuations inevitable at finite temperatures. This performance is evaluated for atoms cooled to $5~\mu$K. In Fig.~\ref{bell}(c), the ensemble-averaged fidelity
$\bar{\mathcal{F}} = {1}/{1001}\sum_{k=1}^{1001}\langle \Phi^+|\rho_k(t)|\Phi^+\rangle$ is also presented (blue solid line), taking into account the Doppler shifts with the thermal velocity spread $\sigma_{v} = 0.022~{\rm m/s}$ from the 780--480~nm two-photon excitation and position fluctuations $(\sigma_{x}, \sigma_{y}, \sigma_{z}) = (22, 25, 60)$~nm~\cite{Chew2022Ultrafast}. When the parameters are initially chosen at the avoided crossings, the fluctuation effectively reduces the chances that the dynamics stay at the resonant point $\rm{K}$. As such, the final fidelity exceeds 99\%, confirming that temperature-induced parameter fluctuations enhance feasibility. For parameters initialized at ${\rm K'}$, such fluctuations leave the Bell-state fidelity essentially unchanged, as shown by the red solid line in Fig.~\ref{bell}(d). This robustness stems from Zeno dynamics, enabling virtual Rydberg bi-excitations that are resilient to fluctuations in Rydberg interactions. Moreover, the numerical analysis presented in Appendix~\ref{app:bbr} demonstrates that the fidelity remains robust against Rydberg-state spontaneous emission and BBR-induced transitions during the coherent excitation stage.

\section{Many-body stabilizer pumping}

This protocol is scalable and allows for stabilizer pumping toward multipartite graph states. The scalability is demonstrated through stabilizer pumping toward the one-dimensional (1D) cluster state, defined as $|C_N\rangle = \frac{1}{2^{N/2}} \sum_{x_1,x_2,\dots,x_N=0}^1 
(-1)^{\sum_{i=1}^{N-1} x_i x_{i+1}} \, |x_1 x_2 \dots x_N\rangle$, which is the unique common eigenstate with eigenvalue $+1$ of the commuting stabilizer generators~\cite{Toth2005Detecting}: $S_1 = X_1 Z_2$, $S_k = Z_{k-1} X_k Z_{k+1}$, $(2 \le k \le N{-}1)$, and $S_N = Z_{N-1} X_N$. These stabilizers, involving tensor products of at most three qubits, are local and therefore naturally compatible with the atomic array geometry shown in Fig.~\ref{fig1}(a). The three-body stabilizer-pumping process follows the same principle as the two-qubit protocol (Fig.~\ref{fig2}). A key distinction, however, is the emergence of an additional dark state, $|G_iG_jG_k\rangle$, which remains frozen despite the laser driving. As shown in Appendix~\ref{sec_steady}, the pumping dynamics still converge to a unique steady state, even when the $-1$ stabilizer eigenspace cannot be completely depleted in a single pumping step. The three-qubit analysis further illustrates how overlapping stabilizer pumps preserve this uniqueness, while the reduced set of direct pumping channels slows convergence.

\begin{figure}
	\centering
	\includegraphics[width=1\linewidth]{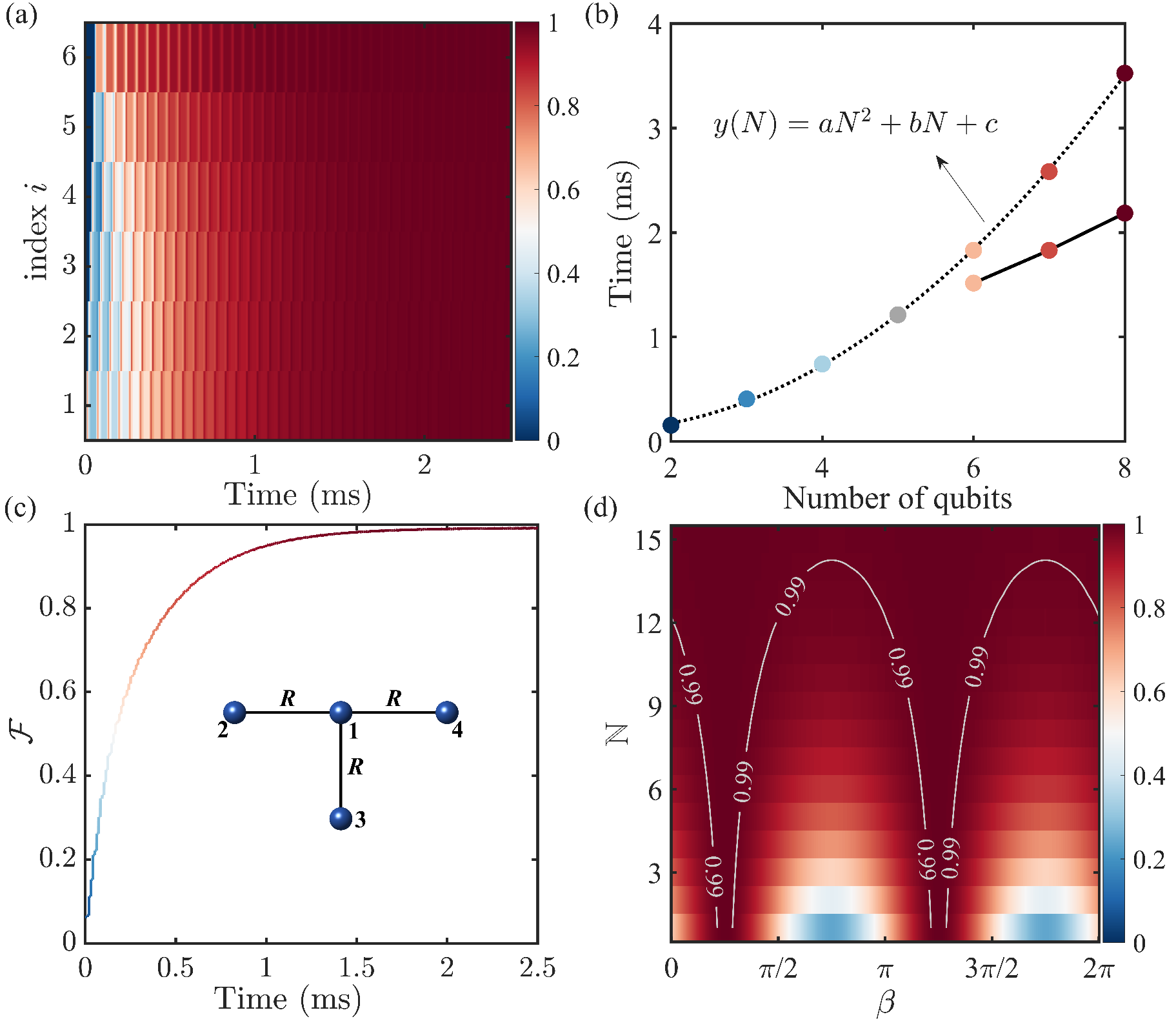}
	\caption{(a) Evolution of stabilizer expectation values $\langle S_i\rangle={\rm Tr}[\rho(t)S_i]$ during stabilizer pumping toward a six-qubit cluster state. 
(b) Convergence time to 99\% fidelity versus system size. The dotted curve is a quadratic fit to the sequential-pumping data ($N=2$--$8$), while the solid polyline connects the parallel-pumping data ($N=6,7,8$) as a guide to the eye. 
(c) Stabilizer pumping toward a four-qubit graph state in a T-shaped geometry. 
(d) GHZ-state purification through stabilizer pumping from perturbed state $|C_3^p\rangle = \cos(\beta) |+0+\rangle + \sin(\beta) |-1-\rangle$.
Parameters: $\Omega_a = 2\pi\times 5~\mathrm{MHz}$, $\Delta=U_{rr}=77\Omega_a$, $\mathcal{N}=100$, $\alpha = 0.6$, $t_b = \mathcal{N}t_a = 0.1~\mu\text{s}$, $\Omega \approx 2\pi\times 13.31~{\rm MHz}$, $\gamma_p = 2\pi\times 6.07~\mathrm{MHz}$, decay duration $t_d = 0.362~\mu{\rm s}$, nearest-neighbor distance $R = 4.7171~\mu{\rm m}$.}\label{fig4}
\end{figure}

The local and composable nature of stabilizer pumping enables the protocol to be extended to many-qubit systems. In Fig.~\ref{fig4}(a), stabilizer pumping toward a six-qubit cluster state is illustrated using the pumping sequence depicted in Fig.~\ref{fig1}(d). Starting from the mixed state, the expectations $\langle S_i \rangle = \mathrm{Tr}[\rho(t) S_i]$ of the six stabilizers converge to an average value of  99.34\% after $\mathbb{N} = 40$ cycles, while the fidelity of the cluster state reaches 99.02\%. This minor difference indicates that the state fidelity cannot in general be inferred from only the $N$ stabilizer-generator expectation values, since an exact evaluation involves all $2^N$ elements of the stabilizer group.

Using an effective model, the convergence time at 99\% fidelity is found to depend quadratically on the system size,  described by $y(N) = a N^2 + b N + c$, with fitted parameters $a = 69.5~\mu{\rm s}$, $b = -139.7~\mu{\rm s}$, and $c = 177.8~\mu{\rm s}$ [dotted line in Fig.~\ref{fig4}(b)]. This scaling is consistent with the known theoretical predictions~\cite{verstraete2009quantum}. 
Remarkably, the convergence time can be substantially reduced by exploiting parallel stabilizer pumping, whereby commuting stabilizers acting on disjoint sets of atoms are driven simultaneously [solid polyline in Fig.~\ref{fig4}(b); see also Appendix~\ref{app:parallel} for details]. These results demonstrate the efficiency and scalability of the stabilizer-pumping scheme. The protocol can in principle be extended to larger systems ($N>8$), while the present numerical calculations are restricted to $N\leq8$ due to the rapidly growing Hilbert space.

The Floquet-Lindblad protocol is readily applicable to alternative two-dimensional geometries. Figure~\ref{fig4}(c) illustrates stabilizer pumping toward a four-qubit graph state---locally equivalent to a Greenberger-Horne-Zeilinger (GHZ) state---in a T-shaped array. The resulting state is stabilized by three two-body operators $\{Z_1 X_2, Z_1 X_3, Z_1 X_4\}$ and one four-body operator $X_1 Z_2 Z_3 Z_4$. The implementation of the latter relies on a selective transition $\ket{-000} \rightarrow \ket{r000}$, achieved by resonantly driving the central qubit (atom 1) while coupling the peripheral qubits (atoms 2--4) via a blue-detuned pulse. This implementation enables deterministic stabilizer-state preparation for MBQC, exemplified by the CNOT gate. More broadly, extensions of the scheme to complete one-dimensional stabilizer pumping, multipartite GHZ states in star-shaped geometries, and two-dimensional cluster states are discussed in Appendix~\ref{app:extensions}.

An important consequence of the attractive pumping dynamics is the purification of imperfect graph states. As shown in Fig.~\ref{fig4}(d), a partially entangled GHZ-like input state $|C_3^p\rangle = \cos(\beta) |+0+\rangle + \sin(\beta) |-1-\rangle$ is driven toward the maximally entangled state $|C_3\rangle = (|+0+\rangle + |-1-\rangle)/\sqrt{2}.$ The purification process requires only a few cycles when $\beta$ is close to $\pi/4$ or $5\pi/4$, indicating rapid convergence near the target manifold. These results show that stabilizer pumping not only facilitates resource state preparation but also provides autonomous purification toward the target stabilizer manifold. Such resilience is particularly relevant for Clifford-based operations in MBQC~\cite{Raussendorf2001A,patil2023clifford, Kumabe2025Complexity}, where maintaining the unmeasured subsystem close to the stabilizer manifold could help mitigate accumulated imperfections.

\section{Conclusion and outlook}

In summary, this work introduces a robust and selective Floquet-engineered stabilizer-pumping scheme for scalable many-body stabilization in neutral atom arrays. Based on the Floquet-Lindblad framework, the approach is inherently resilient to Doppler shifts and spatial fluctuations, and offers a viable pathway to mitigate BBR effects via rapid population recycling in future room-temperature implementations. Its intrinsic purification capability provides autonomous attraction toward target stabilizer manifolds, offering a promising resource for robust MBQC.

To preserve the fidelity established in the first stage of each Floquet cycle, Gaussian pulse shaping is employed in the second stage to suppress the off-resonant crosstalk. This precaution is essential because prolonged interaction time enhances the sensitivity to even minute spectral overlaps. Nevertheless, the protocol remains waveform-flexible: numerical simulations in Appendix~\ref{app:trapezoidal} show that trapezoidal pulses with optimized ramp rates provide sufficient spectral confinement to achieve comparable performance. To address potential atomic losses during prolonged operations, logical state leakage is mitigated by the aforementioned population recycling mechanisms, while trap loss could potentially be reduced by combining our scheme with state-insensitive magic-wavelength tweezers~\cite{Lampen2018Long, Ma2022Universal, Wilson2022Trapping, Bhowmik2024Double, Jansohn2025Magic}.

These results open several near-term opportunities for quantum information processing. The pumping efficiency and robustness may be further enhanced through parallel stabilizer pumping and geometric optimization. Moreover, the compatibility of our protocol with stabilizer-based quantum error-correction codes~\cite{Gong2022Quantum,Zhao2022Realization} suggests a route toward autonomous error mitigation in atomic architectures. Together with recent advances in optical-tweezer control, multi-atom entanglement generation, and dual-beamline architectures that suppress local-addressing crosstalk through high-frequency beat-note averaging~\cite{Radnaev2025Universal}, the present framework provides a resilient path toward large-scale stabilizer-state engineering.
Furthermore, deterministic mapping of atomic cluster states onto optical degrees of freedom, enabled by integration with cavity-QED systems~\cite{ritter2012elementary,thomas2022efficient,wang2025cavity,3fzf-wsr2}, could facilitate efficient entanglement distribution and long-distance quantum communication. Owing to its intrinsic robustness against typical experimental imperfections and its compatibility with existing Rydberg-atom technologies, our framework offers an experimentally accessible platform for near-term validation and a practical tool for autonomous stabilizer-state engineering.

\section*{Acknowledgments}

This work was supported by the National Natural Science Foundation (Grant Nos. 12174048, 12274376, and 12575032), and the Natural Science Foundation of Henan Province under Grant No. 232300421075. W.L. acknowledges support from the EPSRC (Grant No. EP/W015641/1) and the British Council Going Global Partnerships Programme (Contract No. IND/CONT/G/22-23/26).

X.Q.S. conceptualized the study. F.Q.G. drafted the
manuscript and conducted data analysis under the guidance
of S.L.S., W.L., and X.Q.S. All authors contributed to the
review and editing of the manuscript.

\section*{Data availability}

The data supporting the findings of this article are openly available~\cite{Guo2026data}.

\onecolumngrid
\appendix

\section{Derivation of Effective Hamiltonian}\label{sec:eff_Ham}

\subsection{\label{app:zeno}Effective Hamiltonian via Zeno Dynamics}

To obtain stabilizer-selective excitation resistant to Doppler dephasing and position fluctuations, we propose a control sequence that employs unitary kicks~\cite{Facchi2004Unification, Facchi2008Quantum}. 
The system alternates between two driving Hamiltonians over \(\mathcal{N}\) cycles, described by
\begin{eqnarray}\label{eq_fullspace}
H_{\text{total}}(t) =
\begin{cases}
H_{a} \equiv \sum_i \frac{\Omega_a}{2} |r_i\rangle\langle G_i| + \text{H.c.} + H_V, & t \in [nT, nT + t_a), \\
H_{b}(t) \equiv\sum_i \frac{\Omega_b}{2} e^{-i \Delta t} |r_i\rangle\langle G_i| + \text{H.c.} + H_V, & t \in [nT + t_a, nT+T),
\end{cases}
\end{eqnarray}
where $|G_i\rangle$ represents the specific ground state (or qubit state) component coupled to Rydberg state $|r_i\rangle$ by the control fields. The index $i$ denotes the $i$th atom, and \(n = 0, 1, \dots, \mathcal{N}-1\) denotes the cycle index. \(H_V = \sum_{i<j} U_{rr} |r_i\rangle\langle r_i| \otimes |r_j\rangle\langle r_j|\) represents the Rydberg interaction between atoms with strength $U_{rr}=-C_{6}/R^6$.
Each cycle of duration \(T = t_a + t_b\) consists of the system Hamiltonian \(H_{a}\) acting for time \(t_a\) with evolution \(U_{a}(t_a) = \exp(-i H_{a} t_{a})\), followed by the kick Hamiltonian \(H_{b}\) acting for time \(t_b\) with evolution \(U_{b}(t_b) = \mathcal{T}\exp[-i \int_0^{t_b} H_{b}(t) \, dt]\). The total system evolution is given by \(U_{\mathcal{N}}(T) = [U_{b}(t_b) U_{a}(t_a)]^{\mathcal{N}}\). In particular, we can find $i (d/dt_a)\mathcal{V}_{\mathcal{N}}(t_a)=\mathcal{N}H_{\mathcal{N}}(t_a)\mathcal{V}_{\mathcal{N}}(t_a)$ with $\mathcal{V}_{\mathcal{N}}(t_a) = U_{b}^{\dagger \mathcal{N}}(t_b)U_{\mathcal{N}}(T)$, $\mathcal{V}_{\mathcal{N}}(0) = \mathcal{I}$ being the identity, and $H_{\mathcal{N}}(t_a) = 1/\mathcal{N}\sum_{n=0}^{\mathcal{N}-1}U_{b}^{\dagger \mathcal{N}}(U_{b}U_{a})^{n}U_{b}H_{a}U_{b}^{\dagger}(U_{b}U_{a})^{\dagger n}U_{b}^{\mathcal{N}}$. Under the condition $\Omega_b t_b \gg \Omega_a t_a$, we take the large-$\mathcal{N}$ limit while keeping the accumulated evolution time under $H_a$, $\tau=\mathcal{N}t_a$, fixed. The limiting evolution
operator is then given by~\cite{Shao2011Quantum}
\begin{equation}
    \mathcal{U}(\tau)\equiv
    \lim_{\mathcal{N}\to\infty}
    \mathcal{V}_{\mathcal{N}}
    \left(\tau/\mathcal{N}\right),
\end{equation}
satisfying the equation $i\frac{d}{d\tau}\mathcal{U}(\tau)=H_{Z}\mathcal{U}(\tau)$, with $\mathcal{U}(0) = \mathcal{I}$ and ``Zeno" Hamiltonian $H_{Z}\equiv \lim_{\mathcal{N}\to \infty} H_{\mathcal{N}}(\tau/\mathcal{N}) \sim \lim_{\mathcal{N}\to \infty}\frac{1}{\mathcal{N}}\sum_{n=0}^{\mathcal{N}-1}U_{b}^{\dagger n}H_{a}U_{b}^{n}$. Let $U_{b}(t_b) = \exp[-i\sum_{\mu}\lambda_{\mu}(t_b)P_{\mu}(t_b)]$, where $\lambda_{\mu}(t_b)$ is the corresponding real eigen phase and $P_{\mu}(t_b)$ is the corresponding eigen projector associated with the duration $t_b$. Consequently,
\begin{eqnarray}\label{eq:Zeno_Hamiltonian}
    H_{Z} &=& \sum_{\mu}P_{\mu}H_{a}P_{\mu}.
\end{eqnarray}
The evolution operation can be presented as
\begin{eqnarray}\label{eq_UN}
	U_{\mathcal{N}}(T) &\sim& U_{b}^{\mathcal{N}}(t_b)\mathcal{U}(\mathcal{N}t_a)\cr 
	&=& \exp\left[-i\sum_{\mu}\mathcal{N}\lambda_{\mu}P_{\mu}\right] \exp(-iH_{Z}\mathcal{N} t_a)\cr 
	&=& \exp \left[-i\sum_{\mu}\mathcal{N}(\lambda_{\mu}P_{\mu} + P_{\mu}H_{a}P_{\mu} t_a)\right].
\end{eqnarray}

Since \(H_b(t)\) is time-dependent, the Zeno Hamiltonian can be obtained numerically at any given moment from Eq.~(\ref{eq:Zeno_Hamiltonian}), but its analytical expression is generally intractable. To simplify the model, we impose the large-detuning condition and neglect the accompanying Stark shifts. This approximation effectively disregards the special cases where the eigenspectrum of $U_b(t_b)$ exhibits degeneracies near the eigenvalue $+1$, while still capturing the essential physics under typical conditions. Quantitative deviations are evaluated through comparison with full numerical simulations.

\begin{figure}
	\centering
	\includegraphics[width=18cm]{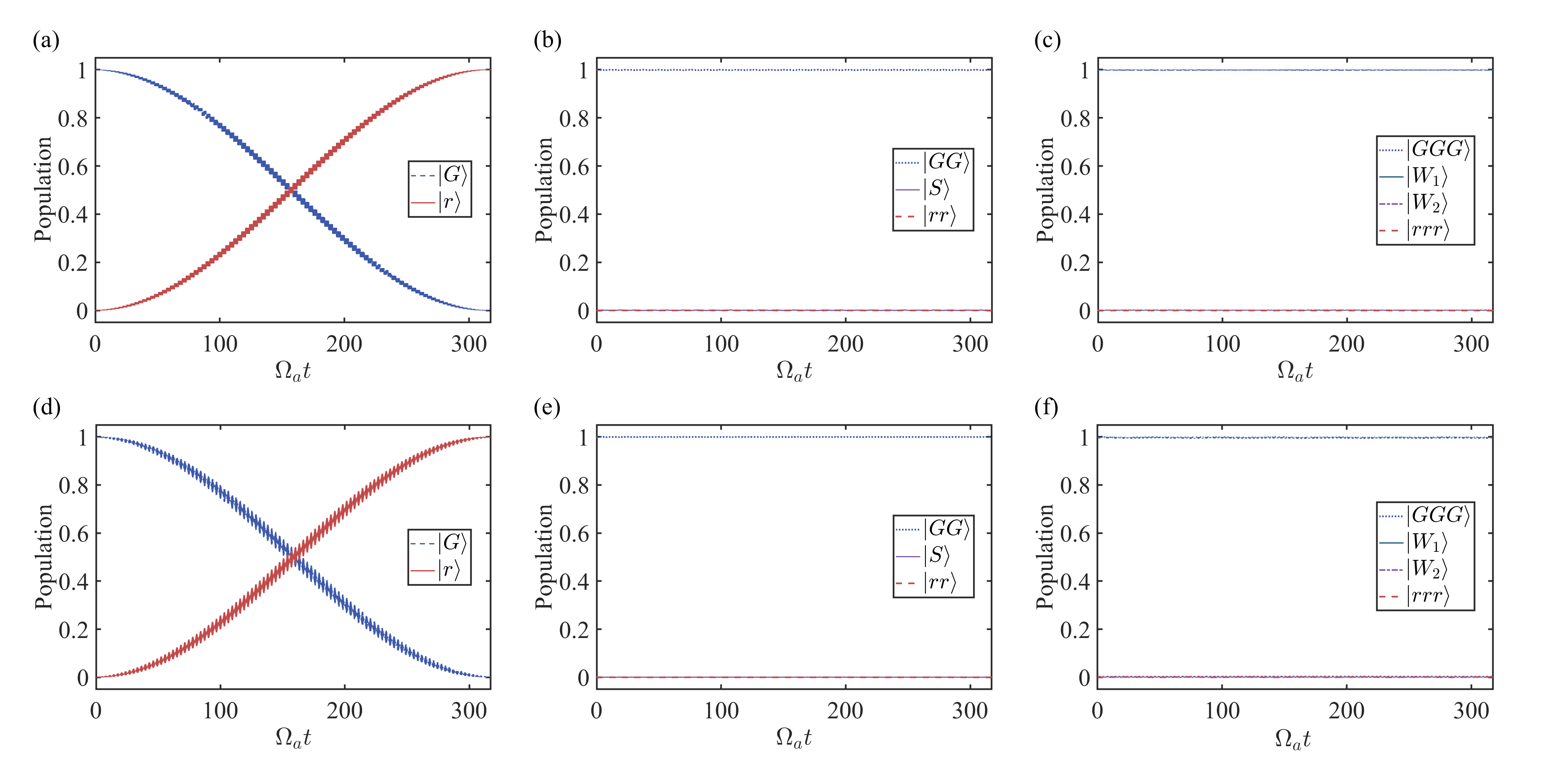}
	\caption{Evolution of basis states with dimensionless parameter $\Omega_a t$ in (a,d) single-, (b,e) two-, and (c,f) three-atom subspaces. (a)-(c) are driven by a constant pulse scheme with $\Delta = U_{rr} = 36\Omega_a$ and $\Omega_b/\Omega_a = 1$ while (d)-(f) are driven by a Gaussian pulse scheme with $\Delta = U_{rr} = 33.68\Omega_a$, $\alpha = 0.6$, and $\Omega/\Omega_a=2.6619$. Other parameters are $\mathcal{N}=100$ and $t_b = \mathcal{N}t_a = \pi/\Omega_a$.}\label{fig_time_evolution}
\end{figure}

Defining the number of atoms coupled with lasers in the subsystem $n_a = \sum_j (|r_j\rangle\langle r_j| + |G_j\rangle\langle G_j|)$, we call $\text{span}\{|GG\rangle, |S\rangle, |rr\rangle \}$ a two-atom subspace ($n_a = 2$) with $|S\rangle = (|Gr\rangle + |rG\rangle)/\sqrt{2}$, $\text{span}\{|GGG\rangle, |W_1\rangle, |W_2\rangle, |rrr\rangle \}$ a three-atom subspace ($n_a = 3$) with $|W_1\rangle = (|GGr\rangle + |GrG\rangle + |rGG\rangle)/\sqrt{3}$, and $|W_2\rangle = (|Grr\rangle + |rGr\rangle + |rrG\rangle)/\sqrt{3}$, and $\text{span}\{|GG^\perp\rangle, |rG^\perp\rangle \}$ (also $\text{span}\{|G^\perp G\rangle, |G^\perp r\rangle \}$, $\text{span}\{|G^\perp G^\perp G\rangle, |G^\perp G^\perp r\rangle,... \}$) a single-atom subspace ($n_a = 1$). Specifically, a single-atom subspace can be simply represented as $\text{span}\{|G\rangle, |r\rangle \}$. For $n_a = 1$, the total Hamiltonian in the single-atom subspace can be rewritten as
\begin{eqnarray}
    H_{a_{1}} &=& \frac{\Omega_{a}}{2}|r\rangle\langle G| + {\rm H.c.}, \cr\cr
    H_{b_{1}} &=& \frac{\Omega_{b}}{2} e^{-i \Delta t} |r\rangle\langle G| + {\rm H.c.}
\end{eqnarray}
In the regime $\Delta \gg \{\Omega_a, \Omega_b\}$, $H_{b_1}\approx0$, the projector from the evolution operation $U_{b_1}$ are $P^{1} = |G\rangle \langle G| + |r\rangle \langle r|$. To describe the evolution of the system over the macroscopic continuous time $t=\mathcal{N}T$, we introduce the time duty cycle factor $t_a/T$, and the corresponding (effective) Zeno Hamiltonian is calculated as
\begin{eqnarray}\label{eq_Z1}
    H_{Z_1} &=& \frac{\Omega_\text{eff}}{2}|r\rangle\langle G| + {\rm H.c.},
\end{eqnarray}
with $\Omega_\text{eff} = \Omega_{a}t_a/T$ an effective Rabi frequency.
Considering $U_{rr} = \Delta$, the total Hamiltonian in the two-atom subspace for $n_a = 2$ can be rewritten as
\begin{eqnarray}
    H_{a_{2}} &=& \frac{\Omega_{a}}{\sqrt{2}}(|S\rangle\langle GG| + |rr\rangle\langle S|) + {\rm H.c.} + \Delta |rr\rangle\langle rr|, \cr\cr
    H_{b_{2}} &=& \frac{\Omega_{b}}{\sqrt{2}} e^{-i \Delta t} (|S\rangle\langle GG| + |rr\rangle\langle S|) + {\rm H.c.} + \Delta |rr\rangle\langle rr|.
\end{eqnarray}
In the regime $\Delta \gg \{\Omega_a, \Omega_b\}$, $H_{b_2}\approx{\Omega_{b}}/{\sqrt{2}}(|S\rangle\langle rr|+|rr\rangle\langle S|)$, the projectors from the evolution operation $U_{b_2}$ are $P^2_{0} = |GG\rangle \langle GG|$, $P^2_{+} = 1/2(|S\rangle\langle S| + |S\rangle\langle rr| + |rr\rangle\langle S| + |rr\rangle\langle rr|)$, and $P^2_- = 1/2(|S\rangle\langle S| - |S\rangle\langle rr| - |rr\rangle\langle S| + |rr\rangle\langle rr|)$. Projected onto the logical ground-state subspace, the corresponding two-atom Zeno Hamiltonian is effectively zero, i.e., 
\begin{eqnarray}\label{eq_Z2}
    H_{Z_2}|GG\rangle &=& 0.
\end{eqnarray}
The total Hamiltonian in the three-atom subspace for $n_a= 3$ can be rewritten as
\begin{eqnarray}
    H_{a_{3}} &=& \frac{\Omega_{a}}{2}(\sqrt{3}|W_1\rangle\langle GGG| + 2|W_2\rangle\langle W_1| + \sqrt{3}|rrr\rangle\langle W_2|) + {\rm H.c.} + \Delta|W_2\rangle\langle W_2| + 3\Delta|rrr\rangle\langle rrr|,\cr\cr
    H_{b_{3}} &=& \frac{\Omega_{b}}{2}e^{-i\Delta t}(\sqrt{3}|W_1\rangle\langle GGG| + 2| W_2\rangle\langle W_1| + \sqrt{3}|rrr\rangle\langle W_2|) + {\rm H.c.} + \Delta|W_2\rangle\langle W_2| + 3\Delta|rrr\rangle\langle rrr|.
\end{eqnarray}
In the regime $\Delta \gg \{\Omega_a, \Omega_b\}$, $H_{b_3}\approx{\Omega_{b}}(|W_1\rangle\langle W_2|+|W_2\rangle\langle W_1|)$, the projectors from the evolution operation $U_{b_3}$ are $P^3_{0} = |GGG\rangle \langle GGG|$, $P^3_{rrr} = |rrr\rangle \langle rrr|$, $P^3_{+} = 1/2(|W_1\rangle\langle W_1| + |W_1\rangle\langle W_2| + |W_2\rangle\langle W_1| + |W_2\rangle\langle W_2|)$, and $P^3_{-} = 1/2(|W_1\rangle\langle W_1| - |W_1\rangle\langle W_2| - |W_2\rangle\langle W_1| + |W_2\rangle\langle W_2|)$. Projected onto the logical ground-state subspace, the corresponding three-atom Zeno Hamiltonian is effectively zero, i.e.,
\begin{eqnarray}\label{eq_Z3}
    H_{Z_3}|GGG\rangle &=& 0.
\end{eqnarray}
As shown in Figs.~\ref{fig_time_evolution}(a)-\ref{fig_time_evolution}(c), we numerically simulate the state evolution of single-, two-, and three-atom subsystems with initial states being ground states ($|G\rangle$, $|GG\rangle$, and $|GGG\rangle$). The single-atom subsystem exhibits coherent Rabi oscillation, whereas excitations from the ground states $|GG\rangle$ and $|GGG\rangle$ in the two- and three-atom subspaces are effectively suppressed. This indicates that multiple atoms undergo only virtual excitation during the system evolution. Consequently, this validates the correctness of our derivations~(\ref{eq_Z1}), (\ref{eq_Z2}), and (\ref{eq_Z3}).

When considering the stabilizer pumping, the ground state $|G_i\rangle$ must be chosen as the local qubit component associated with the stabilizer-violating sector (e.g., one of $|0\rangle$, $|1\rangle$, $|+\rangle$, or $|-\rangle$). Here, we specify the solution required for the main stabilizers. To pump the two-qubit boundary stabilizer $S_{1}=X_{1}Z_{2}$, we require
\begin{eqnarray}
    H_{a}^{S_{1}} &=& \frac{\Omega_{a}}{2}(|r_{1}\rangle\langle -_{1}|\otimes\mathcal{I}_{2} + \mathcal{I}_{1}\otimes|r_{2}\rangle\langle 1_{2}|) + {\rm H.c.} + H_{V},\cr\cr
    H_{b}^{S_{1}}(t) &=& \frac{\Omega_{b}}{2}e^{-i\Delta t}(|r_{1}\rangle\langle -_{1}|\otimes\mathcal{I}_{2} + \mathcal{I}_{1}\otimes|r_{2}\rangle\langle 1_{2}|) + {\rm H.c.} + H_{V}.\nonumber
\end{eqnarray}
The effective Zeno Hamiltonian can be obtained as
\begin{eqnarray}\label{eq:H_S1}
    H_{S_{1}} &=& \frac{\Omega_\text{eff}}{2}(|r_{1}0_{2}\rangle\langle -_{1}0_{2}| + |+_{1}r_{2}\rangle\langle +_{1}1_{2}|) + {\rm H.c.}
\end{eqnarray}
 
To pump the two-qubit boundary stabilizer $S_{N}=Z_{N-1}X_{N}$, we require
\begin{eqnarray}
    H_{a}^{S_{N}} &=& \frac{\Omega_{a}}{2}(|r_{N-1}\rangle\langle 1_{N-1}|\otimes\mathcal{I}_{N} + \mathcal{I}_{N-1}\otimes|r_{N}\rangle\langle -_{N}|) + {\rm H.c.} + H_{V},\cr\cr
    H_{b}^{S_{N}}(t) &=& \frac{\Omega_{b}}{2}e^{-i\Delta t}(|r_{N-1}\rangle\langle 1_{N-1}|\otimes\mathcal{I}_{N} + \mathcal{I}_{N-1}\otimes|r_{N}\rangle\langle -_{N}|) + {\rm H.c.} + H_{V}.\nonumber
\end{eqnarray}
The effective Zeno Hamiltonian can be obtained as
\begin{eqnarray}\label{eq:H_SN}
    H_{S_{N}} &=& \frac{\Omega_\text{eff}}{2}(|r_{N-1}+_{N}\rangle\langle 1_{N-1}+_{N}| + |0_{N-1}r_{N}\rangle\langle 0_{N-1}-_{N}|) + {\rm H.c.}
\end{eqnarray}

To pump the three-qubit bulk stabilizer $S_{k}=Z_{k-1}X_{k}Z_{k+1}~(k=2,3,...,N-1)$, we require
\begin{eqnarray}
    H_{a}^{S_{k}} &=& \frac{\Omega_{a}}{2}(|r_{k-1}\rangle\langle 1_{k-1}|\otimes\mathcal{I}_{k}\otimes\mathcal{I}_{k+1} + \mathcal{I}_{k-1}\otimes|r_{k}\rangle\langle -_{k}|\otimes\mathcal{I}_{k+1} + \mathcal{I}_{k-1}\otimes\mathcal{I}_{k}\otimes|r_{k+1}\rangle\langle 1_{k+1}|) + {\rm H.c.} + H_{V},\cr\cr
    H_{b}^{S_{k}}(t) &=& \frac{\Omega_{b}}{2}e^{-i\Delta t}(|r_{k-1}\rangle\langle 1_{k-1}|\otimes\mathcal{I}_{k}\otimes\mathcal{I}_{k+1} + \mathcal{I}_{k-1}\otimes|r_{k}\rangle\langle -_{k}|\otimes\mathcal{I}_{k+1} + \mathcal{I}_{k-1}\otimes\mathcal{I}_{k}\otimes|r_{k+1}\rangle\langle 1_{k+1}|) + {\rm H.c.} + H_{V}.\nonumber
\end{eqnarray}
The effective Zeno Hamiltonian can be obtained as
\begin{equation}\label{eq:H_Sk}
    H_{S_{k}} = \frac{\Omega_\text{eff}}{2}(|r_{k-1}+_{k}0_{k+1}\rangle\langle 1_{k-1}+_{k}0_{k+1}| + |0_{k-1}r_{k}0_{k+1}\rangle\langle 0_{k-1}-_{k}0_{k+1}| + |0_{k-1}+_{k}r_{k+1}\rangle\langle 0_{k-1}+_{k}1_{k+1}|) + {\rm H.c.}
\end{equation}

\subsection{\label{app:avoided}Explanation of the avoided-crossing phenomenon}

\begin{figure}
	\centering
	\includegraphics[width=18cm]{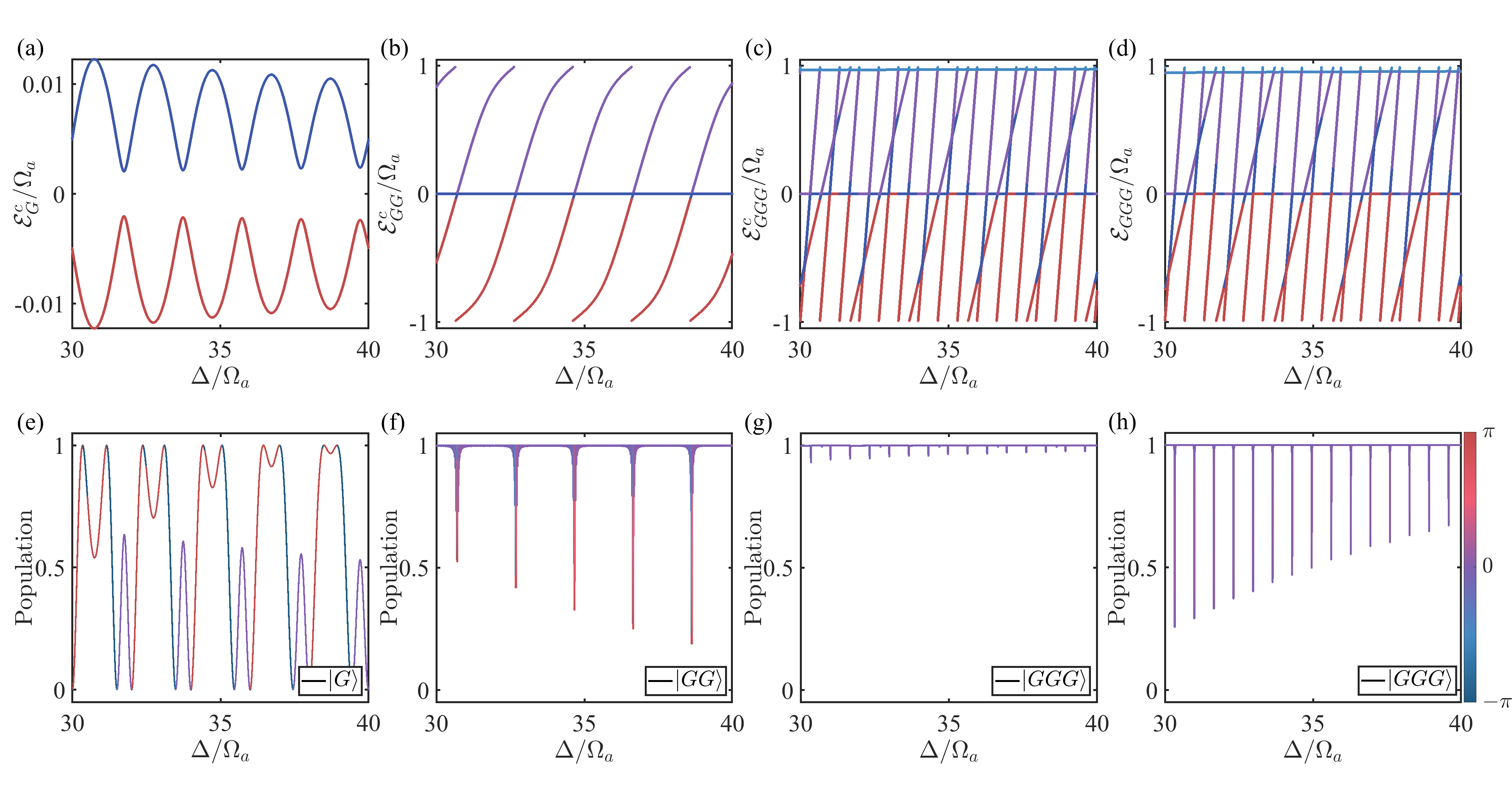}
	\caption{Quasienergy spectra of the Floquet Hamiltonian $H_F$ for single-, two-, and three-atom subsystems under the constant-drive scheme [(a)-(c)] and for three-atom subsystem under the Gaussian-pulse scheme (d), plotted as a function of $\Delta/\Omega_a$. Panels (e)-(h) show the corresponding populations of the basis states for a single atom initialized in $|G\rangle$, two atoms initialized in $|GG\rangle$, and three atoms initialized in $|GGG\rangle$. The colors of the curves in (e)-(h) represent the phase angles of the probability amplitudes.
    Parameters are set as $\Omega_b/\Omega_{a} = 1$ for the constant scheme, and $\Omega/\Omega_{a} = 2.6619$, $\alpha = 0.6$ for the Gaussian pulse scheme, and both have $\mathcal{N} = 100$ and $t_b = \mathcal{N}t_a = \pi/\Omega_a$.}\label{fig_SM_constant}
\end{figure}

The above approximation breaks down when near-degeneracies between the quasienergies of $U_{b}$ result in strong mixing between system states, manifesting as a phenomenon of avoided crossing. Consequently, the effective equations~(\ref{eq_Z1}), (\ref{eq_Z2}), and (\ref{eq_Z3}) are no longer valid in these regions. The exact stroboscopic dynamics are instead described by the Floquet Hamiltonian $H_F = (i/T)\log[U_b(t_b)U_a(t_a)]$. We numerically simulate the corresponding quasienergy spectra and basis-state populations for the single-, two-, and three-atom subsystems, as shown in Fig.~\ref{fig_SM_constant}, where the oscillation in the single-atom case stems from the Stark shifts of Hamiltonian $H_{b_{1}}$, while avoided crossings (corresponding to peaks appearing in the overlap of eigenstates) appear periodically in the quasienergy spectrum. Since \(H_a\) acts only for a short duration within each driving cycle, its contribution to the Floquet quasienergy spectrum is limited to a slight modification of the effective period. The dominant role is instead played by \(H_b(t)\). In the following, we take \(H_b(t)\) in the two-atom model as an example to analytically elucidate the origin of the avoided crossing.

The time-dependent Hamiltonian $H_{b_{2}}$ can be transformed into a time-independent Hamiltonian $H'_{b_{2}}$ by the rotating frame $R_{b_{2}} = \exp[-i\Delta t(|S\rangle\langle S| + 2|rr\rangle\langle rr|)]$
\begin{eqnarray}\label{eq:time-independent Hb2}
    H' _{b_{2}} &=& \frac{\Omega_{b}}{\sqrt{2}}(|S\rangle\langle GG| + |rr\rangle\langle S|)+{\rm H.c.} - \Delta|S\rangle\langle S| - \Delta|rr\rangle\langle rr|.
\end{eqnarray}  
The periodic avoided crossings in the quasienergy spectrum of $U_{b_2}$ are a hallmark of its Floquet dynamics. This phenomenon originates from the periodic ``phase kick" delivered by the rotating-frame transformation $R_{b_2}(t_b) = \mathrm{diag}(1, e^{-i\Delta t_b}, e^{-2i\Delta t_b})$ at each stroboscopic evolution time $t_b$. The nature of the resulting interference is critically dictated by the phase $\phi = \Delta \cdot t_b$ acquired during this period. A strong avoided crossing emerges when this phase kick acts as a symmetry-breaking parity-flip operator, which occurs under the general resonance condition $\Delta \cdot t_b = (2m+1)\pi$ for an integer $m$. Conversely, when the phase kick becomes trivial (occurring at $\Delta \cdot t_b = 2m\pi$), the symmetry of the system is preserved. Between them, there is also a small mixing between the system states due to the existence of systematic nonadiabatic transitions. Notably, this can be avoided by selecting a relatively large detuning in an ideal situation. Furthermore, we provide an analytical argument to demonstrate that the system is primed to exhibit an avoided-crossing phenomenon under the resonance condition $\Delta \cdot t_b = (2m+1) \pi$.

The block matrix describing the high-energy subspace $\{|S\rangle, |rr\rangle\}$ within the time-independent Hamiltonian~(\ref{eq:time-independent Hb2}) is
\begin{equation}\label{eq:HD2}
    H'_D = 
    \begin{pmatrix} 
        -\Delta & \frac{\Omega_{b}}{\sqrt{2}} \\ 
        \frac{\Omega_{b}}{\sqrt{2}} & -\Delta 
    \end{pmatrix},
\end{equation}
which can be diagonalized to yield eigenvalues $E_{\pm} = -\Delta \pm \Omega_{b}/\sqrt{2}$ and eigenstates $|E_{\pm}\rangle = 1/\sqrt{2}(|S\rangle \pm |rr\rangle)$.
We rewrite $H'_{b_{2}}$ in the basis $\{|GG\rangle, |E_+\rangle, |E_-\rangle\}$ as
\begin{equation}
    H'_{S} = \underbrace{\begin{pmatrix} 
        0 & 0 & 0 \\ 
        0 & E_+ & 0 \\ 
        0 & 0 & E_- 
    \end{pmatrix}}_{H_0}
    +
    \underbrace{\begin{pmatrix} 
        0 & \frac{\Omega_{b}}{2} & \frac{\Omega_{b}}{2} \\ 
        \frac{\Omega_{b}}{2} & 0 & 0 \\ 
        \frac{\Omega_{b}}{2} & 0 & 0 
    \end{pmatrix}}_{V'}.
\end{equation}
The transition probability amplitudes for the evolution operator $U_S(t) = \exp(-iH'_S t)$ can be approximated using a Dyson series expansion, since we always assume $\Delta\gg \Omega_b$. To first order, the transition from state $|GG\rangle$ to $|E_{\pm}\rangle$ is given by
\begin{align}
U_{\pm,GG}(t)
\approx -i e^{-iE_\pm t}
\int_0^t e^{iE_\pm t'}
\langle E_\pm|V'|GG\rangle\,dt'
= \frac{\Omega_b}{2}
\frac{e^{-iE_\pm t}-1}{E_\pm}.\end{align}

Transforming back to the $\{|S\rangle, |rr\rangle\}$ basis, we obtain the off-diagonal elements
\begin{align}
    U_{S,GG} &= \frac{1}{\sqrt{2}} [U_{+,GG}(t) + U_{-,GG}(t)] \approx \frac{\Omega_{b}}{2\sqrt{2}} \left( \frac{e^{-iE_+ t} - 1}{E_+} + \frac{e^{-iE_- t} - 1}{E_-} \right), \\
    U_{rr,GG} &= \frac{1}{\sqrt{2}} [U_{+,GG}(t) - U_{-,GG}(t)] \approx \frac{\Omega_{b}}{2\sqrt{2}} \left( \frac{e^{-iE_+ t} - 1}{E_+} - \frac{e^{-iE_- t} - 1}{E_-} \right).
\end{align}
To zeroth order, the elements within the high-energy subspace are
\begin{align}
    U_{S,S} &\approx \frac{1}{2}(e^{-iE_+ t} + e^{-iE_- t}), \quad U_{rr,rr} \approx \frac{1}{2}(e^{-iE_+ t} + e^{-iE_- t}),\\
    U_{S,rr} &\approx \frac{1}{2}(e^{-iE_+ t} - e^{-iE_- t}), \quad U_{rr,S} \approx \frac{1}{2}(e^{-iE_+ t} - e^{-iE_- t}).
\end{align}
Meanwhile, the diagonal element $U_{GG,GG}(t)$ is calculated to the second order:
\begin{eqnarray}
    U_{GG,GG} \approx 1 + U^{(2)}_{GG,GG}(t),
\end{eqnarray}
with 
\begin{eqnarray}
    U^{(2)}_{GG,GG} &=& -\sum_{k \in \{+,-\}} \int_0^t dt_1 \int_0^{t_1} dt_2 \langle GG|V'|E_k\rangle e^{-iE_k(t_1-t_2)}\langle E_k |V'|GG\rangle\cr\cr
    &=& i\frac{\Omega_{b}^2}{4} t \left( \frac{1}{E_+} + \frac{1}{E_-} \right) - \frac{\Omega_{b}^2}{4} \left( \frac{1 - e^{-iE_+ t}}{E_+^2} + \frac{1 - e^{-iE_- t}}{E_-^2} \right).
\end{eqnarray}

The full evolution operator for the kick Hamiltonian is constructed by applying the rotating-frame transformation
\begin{eqnarray}
    U_{b_{2}}(t) &\approx& R_{b_{2}}(t) U'_{b_{2}}(t) \cr\cr
    &=& \begin{pmatrix} 
        1 & 0 & 0 \\ 
        0 & e^{-i \Delta t} & 0\\
        0 & 0 & e^{-2 i \Delta t}  
    \end{pmatrix}
    \begin{pmatrix} 
        U_{GG,GG} & U_{GG,S} & U_{GG,rr} \\ 
        U_{S,GG} & U_{S,S} & U_{S,rr}\\
        U_{rr,GG} & U_{rr,S} & U_{rr,rr}
    \end{pmatrix}.
\end{eqnarray}
Focusing on the high-energy subspace, its dynamics are governed by the effective block matrix $U_{D}$
\begin{eqnarray}
    U_{D}(t) = \begin{pmatrix} 
        e^{-i\Delta t}U_{S,S} & e^{-i\Delta t}U_{S,rr} \\
        e^{-2i\Delta t}U_{rr,S} & e^{-2i\Delta t}U_{rr,rr}
    \end{pmatrix}
    \approx
    \begin{pmatrix} 
        \cos(\frac{\Omega_{b}}{\sqrt{2}} t) & -i \sin(\frac{\Omega_{b}}{\sqrt{2}} t)\\
        -i e^{-i \Delta t}\sin(\frac{\Omega_{b}}{\sqrt{2}}t) & e^{-i \Delta t}\cos(\frac{\Omega_{b}}{\sqrt{2}}t)
    \end{pmatrix}.
\end{eqnarray}
Under the resonance condition $\Delta \cdot t = (2m+1)\pi$, the matrix $U_D$ simplifies and yields eigenvalues of $\pm 1$. The corresponding normalized eigenstates are $|v_1\rangle = -i\cos[(2m+1)\pi\Omega_b /(2\sqrt{2}\Delta)]|S\rangle + \sin[(2m+1)\pi\Omega_b /(2\sqrt{2}\Delta)]|rr\rangle$ for the eigenvalue $+1$, and $|v_2\rangle = i\sin[(2m+1)\pi\Omega_b /(2\sqrt{2}\Delta)]|S\rangle + \cos[(2m+1)\pi\Omega_b /(2\sqrt{2}\Delta)]|rr\rangle$ for the eigenvalue $-1$.
The eigenvalue of $+1$ is of critical importance, as it corresponds to a quasienergy of $\epsilon = 0$ (within the first Brillouin zone). The low-energy ground state $|GG\rangle$ also has a quasienergy approximately equal to zero. Therefore, the resonance condition creates a quasienergy degeneracy: an effective mode $|v_1\rangle$ from the high-energy subspace has nearly the same quasienergy as the ground state $|GG\rangle$. The coupling induces a repulsion between the quasienergy levels, preventing them from crossing, i.e., avoided crossing, which facilitates strong state mixing and resonant population transfer, elucidating the sharp spikes observed in the simulations.

From a more complete numerical perspective taking into account the duration $t_a$ of the resonant transition $H_a$, the condition for avoided crossing to occur during the total period $T = t_a + t_b$ is $\Delta \cdot T = (2m+1)\pi$ in main text.

A similar result also appears in the three-atom ground state Hamiltonian $H_{b_3}$. It can be transformed into 
\begin{eqnarray}\label{eq:time-independent Hb3}
    H'_{b_{3}} &=& \frac{\Omega_{b}}{2}(\sqrt{3}|W_1\rangle\langle GGG| + 2| W_2\rangle\langle W_1| + \sqrt{3}|rrr\rangle\langle W_2|) + {\rm H.c.} - \Delta|W_1\rangle\langle W_1| -\Delta|W_2\rangle\langle W_2|
\end{eqnarray}
by a rotation transformation $R_{b_3} = \exp[-i\Delta t (|W_1\rangle\langle W_1| + 2|W_2\rangle\langle W_2| + 3|rrr\rangle\langle rrr|)]$.

At this point, the block matrix describing the high-energy subspace 
$\{|W_1\rangle, |W_2\rangle\}$ within the time-independent Hamiltonian~(\ref{eq:time-independent Hb3}) 
takes a form similar to that in Eq.~(\ref{eq:HD2}). 
Since the corresponding rotation matrix $R_{b_3}$ contains a detuning that is three times larger, the quasienergy spectrum exhibits not only the same avoided-crossing period as in the two-atom case, but also an additional smaller period equal to one third of the two-atom avoided-crossing period.

\subsection{\label{app:gaussian}Optimization of the scheme}

Figure~\ref{fig_SM_constant}(a) shows a distinct oscillation of the quasienergy spectrum of a single-atom subsystem, which is reflected in the high sensitivity of the state evolution to detuning $\Delta$, as shown in Fig.~\ref{fig_SM_constant}(e). The sharp edges of a square pulse introduce high-frequency components that lead to nonadiabatic excitations, causing the system dynamics to deviate from the idealized theoretical predictions. Therefore, we propose implementing the Floquet-selective drive not with square pulses, but with smooth Gaussian pulses. This approach offers enhanced robustness against parameter noise. The pulse shape is defined as 
\begin{equation}
    \Omega_{b}(t) = \Omega\exp\left[-\frac{(t-2t_{f})^2}{2(\alpha t_{f})^2}\right],
\end{equation}
where \(\Omega\) is the peak amplitude, the total pulse duration is \(t_b = 4t_{f}\), and \(t_f\) is related to the pulse area by \(t_f = \sqrt{\pi/2}/[\alpha \Omega {\rm erf}(\sqrt{2}/\alpha)]\).
We numerically demonstrate the quasienergy spectra of single-, two-, and three-atom subsystems using the Gaussian pulse scheme shown in Figs.~\ref{fig2}(a) and~\ref{fig2}(c) and in Fig.~\ref{fig_SM_constant}(d). Compared with the constant scheme shown in Fig.~\ref{fig_SM_constant}(a), the Gaussian one demonstrates an approximately flat band in the single-atom case that can always induce perfect coherent Rabi oscillations between $|G\rangle$ and $|r\rangle$ at frequency $|\Omega_{\rm eff}|$ (as shown in Figs.~\ref{fig2}(b) and~\ref{fig_time_evolution}(d)), consistent with Eqs.~(\ref{eff}) and~(\ref{eq_Z1}). In addition, we demonstrate the temporal evolution of two- and three-atom subsystems (Figs.~\ref{fig_time_evolution}(e) and~\ref{fig_time_evolution}(f)), as well as the changes in detuning $\Delta$ (Figs.~\ref{fig_SM_constant}(d) and~\ref{fig_SM_constant}(h)), which exhibit behavior similar to that of the constant-pulse scheme.

\begin{figure}
	\centering
	\includegraphics[width=18cm]{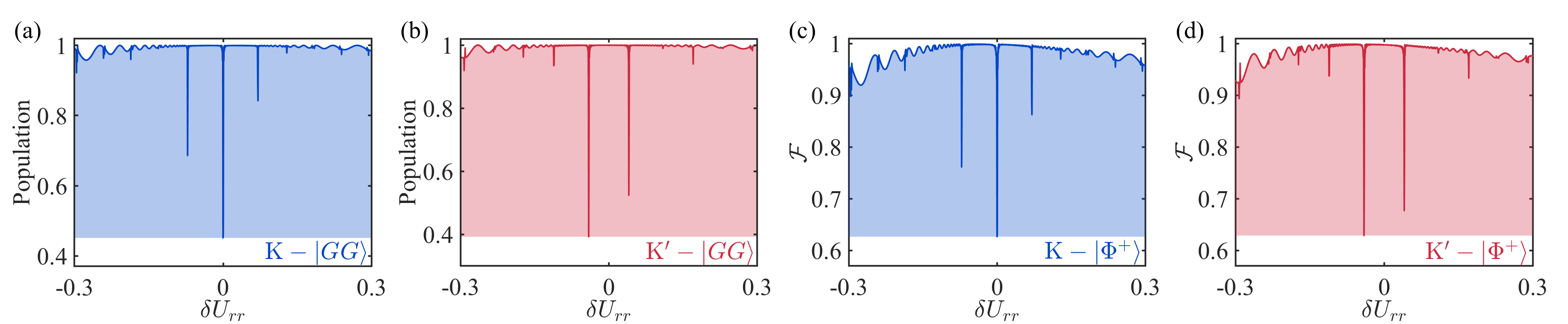}
	\caption{Population of $|GG\rangle$ [(a), (b)] and fidelity of $|\Phi^+\rangle$ [(c), (d)] as a function of the static error in the Rydberg interaction, $\delta U_{rr} \in [-0.3, 0.3]$. The perturbed interaction is modeled as $U_{rr}(1 + \delta U_{rr})$. Panels (a) and (c) correspond to the operating point $\mathrm{K}$, characterized by $\Delta = U_{rr} = 36.655\Omega_a$ and an interatomic distance of $R = 5.3382\,\mu\mathrm{m}$. Panels (b) and (d) correspond to point $\mathrm{K}'$ with $\Delta = U_{rr} = 33.68\Omega_a$ and $R = 5.4141\,\mu\mathrm{m}$. Other parameters are $\Omega_a/(2\pi) = 5\,\mathrm{MHz}$, $\mathcal{N}=100$, $\alpha = 0.6$, $t_b = \mathcal{N}t_a = 0.1\,\mu\mathrm{s}$, $\mathbb{N} = 10$, $\Omega \approx 2\pi\times 13.31\,\mathrm{MHz}$, and decay duration $t_d = 0.362\,\mu\mathrm{s}$.}
\label{fixed_fluctuation}
\end{figure}

Finally, we attribute the robustness of the stabilizer-pumping scheme against parameter fluctuations to the underlying quantum Zeno dynamics. This is verified by introducing a static error scaling to the Rydberg interaction, given by $U_{rr} (1 +\delta U_{rr})$. Figure~\ref{fixed_fluctuation} displays the population of $|GG\rangle$ state and fidelity of Bell state $|\Phi^+\rangle$ across the range $\delta U_{rr} \in [-0.3, 0.3]$, demonstrating that high fidelity is sustained throughout this broad interval, except in narrow resonance regions.

\subsection{\label{app:trapezoidal}Trapezoidal-pulse implementation}

Maintaining a strong average Rabi frequency with Gaussian pulses typically requires relatively high laser power. To mitigate this, we can also consider trapezoidal pulses, which are widely used in experiments~\cite{Ebadi2021Quantum, PhysRevX.14.011025} and are particularly suitable for our scheme. Accordingly, in the second stage of Floquet-selective excitation discussed in the main text, the Gaussian pulse can be replaced by a trapezoidal pulse defined by
\begin{eqnarray}
    \Omega_b(t) = 
\begin{cases} 
\Omega_{\rm op} t/t_{\text{rise}} & 0 \le t < t_{\text{rise}} \\
\Omega_{\rm op} & t_{\text{rise}} \le t \le t_b - t_{\text{rise}} \\
\Omega_{\rm op} (t_b - t)/t_{\text{rise}} & t_b - t_{\text{rise}} < t \le t_b
\end{cases}
\end{eqnarray}
where the peak Rabi frequency is $\max(\Omega_b)=\Omega_{\rm op} = \Omega_a = 2\pi \times 5~\mathrm{MHz}$, the rise time is $t_{\text{rise}} = r_{\rm ramp} t_b / 2$ with the ramp fraction $r_{\rm ramp} = 1/3$, and the pulse area is constrained by $\int_0^{t_b} \Omega_b(t)dt = 1.2 \pi$.

\begin{figure}
	\centering
	\includegraphics[width=1\linewidth]{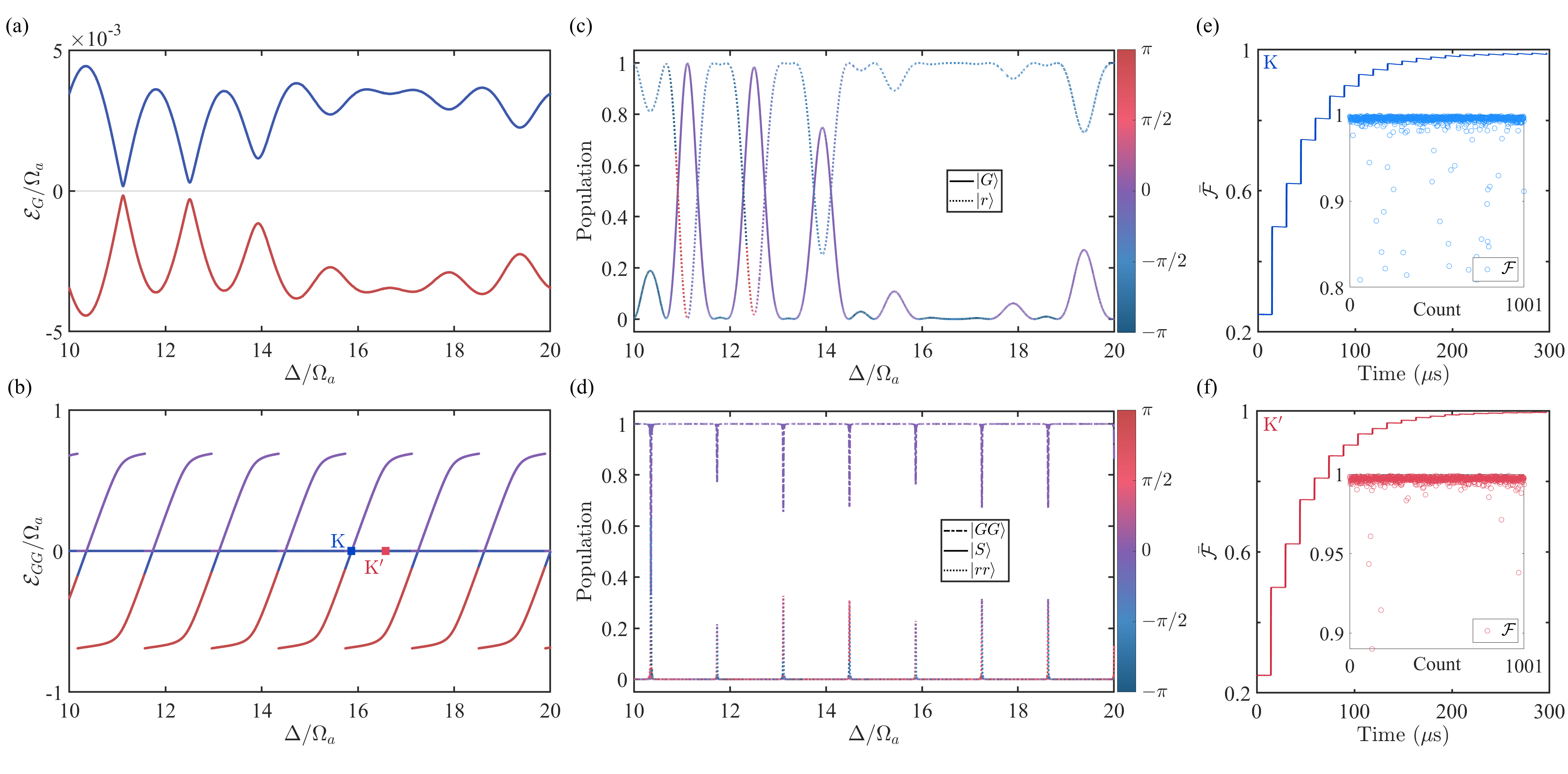}
	\caption{(a) Quasienergy of the effective single-atom Floquet Hamiltonian and (c) populations for a single atom initialized in \(|G\rangle\) versus detuning $\Delta~(=U_{rr})$. 
(b, d) Quasienergy \(\mathcal{E}_{GG}\) and populations for the two-atom system initialized in \(|GG\rangle\). Curve colors in (c, d) represent phase angles of probability amplitudes associated with the corresponding quantum states.
(e, f) Bell state fidelity at avoided-crossing point ${\rm K}$ (\(\Delta/\Omega_a = 15.871\)) and regular point ${\rm K'}$ (\(\Delta/\Omega_a = 16.5597\)). Simulations at $5~\mu{\rm K}$ are averaged over 1001 random spatial realizations (insets). 
Points ${\rm K}$ and ${\rm K'}$ correspond to $R = 6.1376~\mu\text{m}$ and $6.0942~\mu\text{m}$. Other parameters: $\Omega_{\rm op}=\Omega_a = 2\pi\times 5~\mathrm{MHz}$, $\mathcal{N}=100$, $t_a = 1~\text{ns}$, $t_b = 0.144~\mu\text{s}$, $\mathbb{N} = 10$, $\gamma_p = 2\pi\times 6.07~\mathrm{MHz}$, $t_d = 0.362~\mu{\rm s}$.}\label{fig_Trapezoidal}
\end{figure}

Figures~\ref{fig_Trapezoidal}(a)-\ref{fig_Trapezoidal}(d) show the resulting quasienergy spectra and population dynamics in the effective single-atom and two-atom subspaces. In the two-atom subspace, the trapezoidal-pulse scheme achieves performance comparable to that of the Gaussian-pulse scheme, while in the effective single-atom subspace, the populations of $|G\rangle$ and $|r\rangle$ vary more noticeably with detuning $\Delta$, reflecting the oscillations in the quasienergy spectrum. Following the same procedure as before, we select two operational points for Bell-state stabilizer pumping: the avoided crossing $\mathrm{K}$ at $\Delta/\Omega_a = 15.871$ and the regular point $\mathrm{K}'$ at $\Delta/\Omega_a = 16.5597$. As illustrated in Figs.~\ref{fig_Trapezoidal}(e) and \ref{fig_Trapezoidal}(f), both points maintain high-fidelity performance.

\section{Engineered dissipative reset from the Rydberg state}

\subsection{\label{app:reset}Fast dissipative reset}

The lifetime of the Rydberg state $|r\rangle$ is often on the order of hundreds of microseconds, which is much longer than that of the intermediate low-lying excited state $\ket{p}$ (typically a few tens of nanoseconds). For simplicity, we neglect the spontaneous emission rate $\gamma_{r}$ of the Rydberg state in this stage.
 In addition, the intermediate state is assumed to decay exclusively to the sector $|0\rangle$ due to the selection rule (this can be realized by choosing the stretched $nP$ states, for example, $|p\rangle \equiv |5P_{3/2}, F=3, m_{F}=3\rangle$ with a lifetime $\tau_{p} = 1/\gamma_{p} = 26.2$ ns). By utilizing an additional laser to couple resonantly $|r \rangle$ to $|p\rangle$ with a strength of $\Omega_{p}$, we can realize the engineered dissipative reset to the ground state $|0\rangle$.

From the perspective of Liouvillian space, the Liouvillian eigenvalue spectrum determines the timescale of system relaxation to stationarity~\cite{Carollo2021Exponentially, Zhou2023Accelerating}. This can be used to modulate the relaxation time; for instance, by tuning the initial subspace, the relaxation rate can be effectively controlled, reminiscent of the Mpemba effect in quantum systems. For a three-level system spanned by $|0\rangle \equiv (1~0~0)^{T}$, $|p\rangle \equiv (0~1~0)^{T}$, and $|r\rangle \equiv (0~0~1)^{T}$, subject to the driving Hamiltonian $H_{rp} = \Omega_{p}/2|r\rangle\langle p| + {\rm H.c.}$, the Liouvillian superoperator is given by
\begin{eqnarray}
    \mathcal{L}_{d} = -i(H_{rp}\otimes \mathcal{I} - \mathcal{I}\otimes H_{rp}^{\rm T}) + L_{0p}\otimes L_{0p}^{*} - \frac{1}{2}(L_{0p}^{\dagger}L_{0p}\otimes \mathcal{I} + \mathcal{I} \otimes L_{0p}^{\rm T}L_{0p}^{*})
\end{eqnarray}
with $L_{0p} = \sqrt{\gamma_{p}}|0\rangle\langle p|$. The eigenvalues are 
\[\lambda_{0} = 0,~~~\lambda_{1,2} = -(\gamma_{p} - \kappa)/4,~~~\lambda_{3,4} = -(\gamma_{p} + \kappa)/4,~~~\lambda_{5} = -(\gamma_{p} - \kappa)/2,~~~\lambda_{6,7} = -\gamma_{p}/2,~~~\lambda_{8} = -(\gamma_{p} + \kappa)/2,\]
with $\kappa = \sqrt{\gamma_{p}^2 - 4\Omega_{p}^2}$, the corresponding right and left eigenmatrices are 
\begin{align}
r_0 &= \begin{pmatrix} 1 & 0 & 0 \\ 0 & 0 & 0 \\ 0 & 0 & 0 \end{pmatrix}, \quad
r_1 = \frac{1}{N_{-}} \begin{pmatrix} 0 & 0 & 0 \\ -i (\gmp - \kpp) & 0 & 0 \\ 2 \omp & 0 & 0 \end{pmatrix}, \quad
r_2 = \frac{1}{N_{-}} \begin{pmatrix} 0 & i (\gmp - \kpp) & 2 \omp \\ 0 & 0 & 0 \\ 0 & 0 & 0 \end{pmatrix}, \cr\cr
r_3 &= \frac{1}{N_{+}} \begin{pmatrix} 0 & 0 & 0 \\ -i (\gmp + \kpp) & 0 & 0 \\ 2 \omp & 0 & 0 \end{pmatrix}, \quad
r_4 = \frac{1}{N_{+}} \begin{pmatrix} 0 & i (\gmp + \kpp) & 2 \omp \\ 0 & 0 & 0 \\ 0 & 0 & 0 \end{pmatrix}, \quad
r_5 = \begin{pmatrix} -\frac{\gmp (\gmp - \kpp)}{2 \omp^2} & 0 & 0 \\ 0 & 1 - \frac{\kpp (\gmp - \kpp)}{2 \omp^2} & -\frac{i (\gmp - \kpp)}{2 \omp} \\ 0 & \frac{i (\gmp - \kpp)}{2 \omp} & 1 \end{pmatrix}, \cr\cr
r_6 &= \frac{\sqrt{2}}{D_0} \begin{pmatrix} -2 \omp & 0 & 0 \\ 0 & \omp & -i \gmp/2 \\ 0 & i \gmp/2 & \omp \end{pmatrix}, \quad
r_7 = \frac{1}{\sqrt{2}} \begin{pmatrix} 0 & 0 & 0 \\ 0 & 0 & 1 \\ 0 & 1 & 0 \end{pmatrix}, \quad
r_8 = \begin{pmatrix} -\frac{\gmp (\gmp + \kpp)}{2 \omp^2} & 0 & 0 \\ 0 & -1 + \frac{\gmp (\gmp + \kpp)}{2 \omp^2} & -\frac{i (\gmp + \kpp)}{2 \omp} \\ 0 & \frac{i (\gmp + \kpp)}{2 \omp} & 1 \end{pmatrix},\nonumber
\end{align}
and
\begin{align}
l_0^{\dagger} &= \begin{pmatrix} 1 & 0 & 0 \\ 0 & 1 & 0 \\ 0 & 0 & 1 \end{pmatrix}, \quad
l_1^{\dagger} = \frac{N_{-}}{4\omp\kpp} \begin{pmatrix} 0 & -2i\omp & \gmp + \kpp \\ 0 & 0 & 0 \\ 0 & 0 & 0 \end{pmatrix}, \quad
l_2^{\dagger} = \frac{N_{-}}{4\omp\kpp} \begin{pmatrix} 0 & 0 & 0 \\ 2i\omp & 0 & 0 \\ \gmp + \kpp & 0 & 0 \end{pmatrix},  \cr\cr
l_3^{\dagger} &= \frac{N_{+}}{4\omp\kpp} \begin{pmatrix} 0 & 2i\omp & -\gmp + \kpp \\ 0 & 0 & 0 \\ 0 & 0 & 0 \end{pmatrix}, \quad
l_4^{\dagger} = \frac{N_{+}}{4\omp\kpp} \begin{pmatrix} 0 & 0 & 0 \\ -2i\omp & 0 & 0 \\ -\gmp + \kpp & 0 & 0 \end{pmatrix}, \quad
l_5^{\dagger} = \frac{1}{4\kpp^2}
\begin{pmatrix}
0 & 0 & 0 \\
0 & 4\Omega_p^2 & 2i\omp(\gmp+\kpp) \\
0 & -2i\omp(\gmp+\kpp) & (\gmp+\kpp)^2
\end{pmatrix}, \cr\cr
l_6^{\dagger} &= -\frac{\sqrt{2}D_0}{\kpp^2} \begin{pmatrix} 0 & 0 & 0 \\ 0 & \omp & i\gmp/2 \\ 0 & -i\gmp/2 & \omp \end{pmatrix}, \quad
l_7^{\dagger} = \frac{1}{\sqrt{2}} \begin{pmatrix} 0 & 0 & 0 \\ 0 & 0 & 1 \\ 0 & 1 & 0 \end{pmatrix}, \quad
l_8^{\dagger} = \frac{1}{4\kpp^2}
\begin{pmatrix}
0 & 0 & 0 \\
0 & 4\Omega_p^2 & 2i\omp(\gmp-\kpp) \\
0 & -2i\omp(\gmp-\kpp) & (\gmp-\kpp)^2
\end{pmatrix},\nonumber
\end{align}
with $N_{\pm} = \sqrt{4\omp^2 + |\gmp \pm \kpp|^2}$ and $D_0 = \sqrt{16\omp^2 + \kpp^2}$.

This system exhibits a third-order ($\lambda_{5} = \lambda_{6} = \lambda_{8}, r_5\propto r_6\propto r_8$) and two second-order LEPs ($\lambda_{1} = \lambda_{3}, r_{1} = r_{3}$ and $\lambda_{2} = \lambda_{4}, r_{2} = r_{4}$) when $\Omega_{p} = 0.5\gamma_{p}$.

\begin{figure}
	\centering
	\includegraphics[width=1\linewidth]{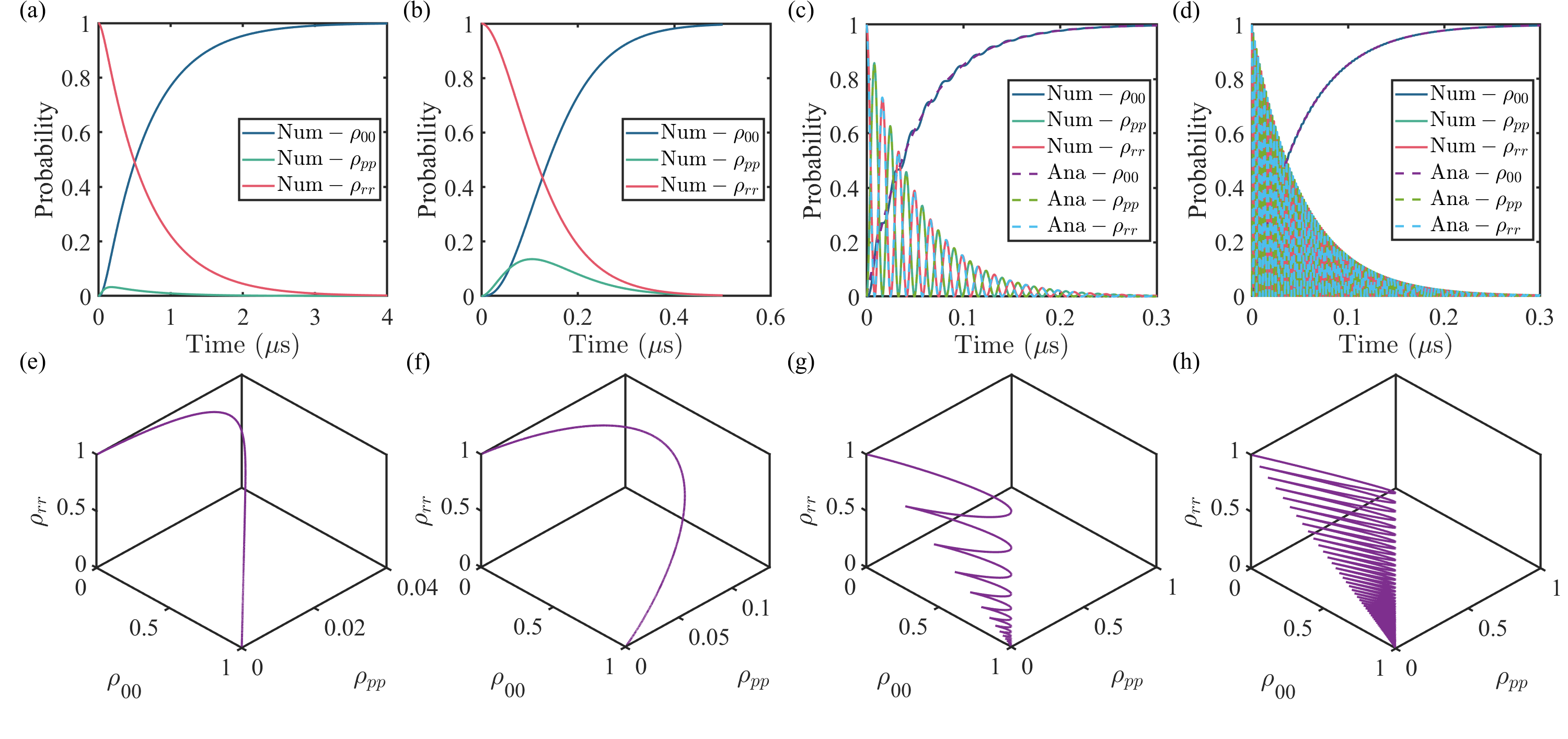}
	\caption{Probabilities of $\rho_{00}$, $\rho_{pp}$, $\rho_{rr}$ for the conditions (a) $\Omega_{p} = 0.2\gamma_{p}$; (b) $\Omega_{p} = 0.5\gamma_{p}$; (c) $\Omega_{p} = 10\gamma_{p}$; (d) $\Omega_{p} = 50\gamma_{p}$ with a single atom initialized in $|r\rangle$. (c) and (d) additionally consider analytical solutions in Eq.~(\ref{fig_outweigh}) when $\Omega_{p}\gg \gamma_{p}$, showing close agreement between numerical and analytical results. (e)-(h) show the relaxation process for three states $\rho_{00}$, $\rho_{pp}$, $\rho_{rr}$ denoted by probabilities, ultimately reaching stability. The decay rate of the intermediate state $|p\rangle$ is $\gamma_p/(2\pi) = 6.07~{\rm MHz}$.}\label{fig_relaxation}
\end{figure}

We propose an alternative acceleration method valid for $\Omega_{p} > 0.5\gamma_{p}$ when considering the system relaxation to the steady state $\rho_{ss}$. As shown in Fig.~\ref{fig_relaxation}, the system relaxes progressively faster as $\Omega_p$ exceeds $\gamma_p$, and reaches a stable relaxation regime ($\Omega_{p} \gg \gamma_{p}$) with a maximum relaxation rate of $\gamma_{p}/2$. Next, we provide an analytical process. 

Assuming the initial state $\rho(0) = \rho_{rr}$, we thus have ${\rm Tr}[l_{1-4}^\dagger\rho_{rr}]={\rm Tr}[l_{7}^\dagger\rho_{rr}]=0$. 
For this specific initial state $\rho_{rr}$, the relaxation is governed by the slowest populated mode $\lambda_{5}$. We thus reduce the evolution state into the time-dependent density matrix 
\begin{eqnarray}
    \rho(t) &=& r_{0} + c_{5}(t)r_{5} + c_{6}(t)r_{6} + c_{8}(t)r_{8}\cr\cr
    &=& r_{0} + \frac{\kappa (\gamma_{p} + \kappa) + 2 \Omega_{p}^2}{2 \kappa^2}e^{\lambda_{5}t} r_{5} - \frac{\sqrt{2} \Omega_{p} \sqrt{16 \Omega_{p}^2 + \kappa^2}}{\kappa^2}e^{\lambda_{6}t} r_{6} + \frac{\kappa (\kappa - \gamma_{p}) + 2 \Omega_{p}^2}{2 \kappa^2}e^{\lambda_{8}t} r_{8} \cr\cr
    &=& \left\{1 + \frac{e^{-\frac{\gamma_{p}t}{2}} \left[ 4 \Omega_{p}^2 - \left( \kappa^2 + 4 \Omega_{p}^2 \right) \cosh\left( \frac{t \kappa}{2} \right) - \gamma_{p} \kappa \sinh\left( \frac{t \kappa}{2} \right) \right]}{\kappa^2}\right\}|0\rangle\langle 0|
    + \frac{4 e^{-\frac{\gamma_{p}t}{2}} \Omega_{p}^2 \sinh^2\left( \frac{t \kappa}{4} \right)}{\kappa^2}|p\rangle\langle p|\cr\cr
    &&- \frac{i e^{-\frac{\gamma_{p}t}{2}} \Omega_{p} \left[ 2 \gamma_{p} \sinh^2\left( \frac{t \kappa}{4} \right) + \kappa \sinh\left( \frac{t \kappa}{2} \right) \right]}{\kappa^2}|p\rangle\langle r|
    + \frac{i e^{-\frac{\gamma_{p}t}{2}} \Omega_{p} \left[ 2 \gamma_{p} \sinh^2\left( \frac{t \kappa}{4} \right) + \kappa \sinh\left( \frac{t \kappa}{2} \right) \right]}{\kappa^2}|r\rangle\langle p|\cr\cr
    &&+ \frac{e^{-\frac{\gamma_{p}t}{2}} \left[ -2 \Omega_{p}^2 + \left( \kappa^2 + 2 \Omega_{p}^2 \right) \cosh\left( \frac{t \kappa}{2} \right) + \gamma_{p} \kappa \sinh\left( \frac{t \kappa}{2} \right) \right]}{\kappa^2}|r\rangle\langle r|.
\end{eqnarray}

As shown in Figs.~\ref{fig_relaxation}(c) and~\ref{fig_relaxation}(d), this can be numerically simulated to obtain consistent results. Specifically, the coherence $\rho_{rp}+\rho_{pr}$ does not contribute to the dynamics. Furthermore, we observe that the relaxation of $\rho_{00}(t)$ accelerates with the increase in the exchange rate between $\rho_{pp}(t)$ and $\rho_{rr}(t)$, as shown in Figs.~\ref{fig_relaxation}(e)-\ref{fig_relaxation}(h). 
In detail, $\rho_{00}(t)$ can also be rewritten as
\begin{eqnarray}
    \rho_{00}(t) &=& 1 - \rho_{pp}(t) - \rho_{rr}(t)\cr\cr
    &=& 1 + \frac{e^{-\frac{\gamma_{p}t}{2}} \left[ 4 \Omega_{p}^2 - \left( \kappa^2 + 4 \Omega_{p}^2 \right) \cosh\left( \frac{t \kappa}{2} \right) - \gamma_{p} \kappa \sinh\left( \frac{t \kappa}{2} \right) \right]}{\kappa^2}\cr\cr
    &=& 1 - e^{-\frac{\gamma_{p}t}{2}} + e^{-\frac{\gamma_{p}t}{2}} \cdot \frac{\gamma_{p}^2[1-\cosh(\frac{\kappa t}{2})]-\gamma_{p}\kappa \sinh(\frac{\kappa t}{2})}{\kappa^2}.
\end{eqnarray}
The value of $\kappa^2$ determines three different mechanisms of the system: (i) $\kappa^2>0$ corresponds to an overdamped mechanism. We rewrite
\begin{eqnarray}
    \rho_{00}(t) = 1 - \frac{\gamma_{p}^2+\gamma_{p} \kappa}{2\kappa^2}e^{\lambda_{5}t} - (1-\frac{\gamma_{p}^2}{\kappa^2})e^{\lambda_{6}t} - \frac{\gamma_{p}^2-\gamma_{p} \kappa}{2\kappa^2}e^{\lambda_{8}t}.
\end{eqnarray}
It has a slower decay mode $\lambda_{5} = -(\gamma_{p} - \kappa)/2$, a mode $\lambda_{6} = -\gamma_{p}/2$, and a faster decay mode $\lambda_{8} = -(\gamma_{p} + \kappa)/2$. At this point, the relaxation rate of the system is mainly determined by the slower mode $\lambda_{5}$. Under extreme conditions $\Omega_{p} \ll \gamma_{p}$, we further expand $\Omega_{p}$ to second order 
\begin{eqnarray}
    \rho_{00}(t) \approx 1 - \exp\left({-\frac{\Omega_{p}^2t}{\gamma_{p}}}\right).
\end{eqnarray}
The result is the same as that obtained by employing perturbation theory and adiabatic elimination.

(ii) $\kappa^2=0$ corresponds to the critical damping mechanism. The general eigenmode expression becomes singular at the LEP, but a finite analytical limit can be obtained as $\kappa\to0$. This can be achieved through the use of L$^\prime$H$\rm \hat{o}$pital$^\prime$s rule, deriving 
\begin{eqnarray}
    \rho_{00}(t) &=& \lim_{\kappa \to 0} 
    1 - e^{-\frac{\gamma_{p}t}{2}} + e^{-\frac{\gamma_{p}t}{2}} \cdot \frac{\gamma_{p}^2[1-\cosh(\frac{\kappa t}{2})]-\gamma_{p}\kappa \sinh(\frac{\kappa t}{2})}{\kappa^2}\cr\cr
    &=& 1-\exp\left[-\frac{\gamma_pt}{2}+\ln\left(1+ \frac{\gamma_{p} t}{2} + \frac{\gamma_{p}^2 t^2}{8}\right)\right].
\end{eqnarray}

(iii) $\kappa^2<0$ corresponds to an underdamped mechanism. Unlike the first two mechanisms, the system experiences strong oscillations and eventually reaches a steady state. However, we propose that when $\Omega_{p}\gg\gamma_{p}$, the system will exhibit an approximate critical damping phenomenon under the underdamped mechanism presented by
\begin{eqnarray}
    \rho_{00}(t) &\approx& 1 -\exp\left({-\frac{\gamma_{p} t}{2}}\right).
\end{eqnarray}
At this point, the system exhibits faster relaxation than at the LEP. In addition, the effective form of the entire system can be demonstrated as
\begin{eqnarray}\label{fig_outweigh}
    \rho(t) &\approx&
(1 - e^{-\frac{\gamma_{p} t}{2}})|0\rangle\langle 0|
+ \frac{1 - \cos (\Omega_{p} t)}{2}e^{-\frac{\gamma_{p} t}{2}} |p\rangle\langle p|
- \frac{i \sin (\Omega_{p} t)}{2} e^{-\frac{\gamma_{p} t}{2}} |p\rangle\langle r|
\cr\cr
&&+\frac{i \sin (\Omega_{p} t)}{2} e^{-\frac{\gamma_{p} t}{2}}|r\rangle\langle p|
+ \frac{1 + \cos (\Omega_{p} t)}{2}e^{-\frac{\gamma_{p} t}{2}}|r\rangle\langle r|.
\end{eqnarray}
Meanwhile, it can be observed that the evolution of the states $|p\rangle$ and $|r\rangle$ still maintains high-speed oscillations with $\Omega_{p}$ as the frequency.

\subsection{\label{app:bbr}Mitigation of BBR-induced leakage}

As operation durations extend into the hundreds of microseconds, room-temperature BBR errors become a significant concern. As a potential extension of our protocol, we provide a strategy to mitigate these BBR-induced effects.
The system dynamics during the coherent excitation stage are governed by the Lindblad master equation
\begin{align}
   \dot{\rho}(t) &= -i[H_{\chi},\rho(t)] \nonumber \\
   &\quad + \sum_{n,k} \left( L_{kr}^{(n)}\rho(t) L_{kr}^{(n)^\dagger} - \frac{1}{2}\left\{L_{kr}^{(n)^\dagger}L_{kr}^{(n)}, \rho(t)\right\} \right) \nonumber \\
   &\quad + \sum_{n} \left( L_{0p}^{(n)}\rho(t) L_{0p}^{(n)^\dagger} - \frac{1}{2}\left\{L_{0p}^{(n)^\dagger}L_{0p}^{(n)}, \rho(t)\right\} \right),
\end{align}
where $n$ indexes the atoms and $k$ denotes the decay channels associated with the Rydberg state $|r\rangle$. We account for spontaneous emission from $|r\rangle$ via two channels, defined by the jump operators $L_{0r}^{(n)} = \sqrt{\gamma_{r}/2}|0_{n}\rangle\langle r_{n}|$ and $L_{1r}^{(n)} = \sqrt{\gamma_{r}/2}|1_{n}\rangle\langle r_{n}|$, with a decay rate $\gamma_{r}/(2\pi) = 0.28~\mathrm{kHz}$. Additionally, the decay from the intermediate state $|p\rangle$ is described by $L_{0p}^{(n)} = \sqrt{\gamma_{p}}|0_{n}\rangle\langle p_{n}|$ with $\gamma_p/(2\pi) = 6.07~\mathrm{MHz}$. 

To model BBR at a temperature of $300~\mathrm{K}$~\cite{SIBALIC2017319}, we include decays to the neighboring Rydberg states $|R^{1}\rangle \equiv |79P_{1/2}\rangle$, $|R^{2}\rangle \equiv |79P_{3/2}\rangle$, $|R^{3}\rangle \equiv |80P_{1/2}\rangle$, and $|R^{4}\rangle \equiv |80P_{3/2}\rangle$. The corresponding Lindblad operators are given by $L_{R^{1}r}^{(n)} = \sqrt{0.17\gamma_{\mathrm{BBR}}}|R^{1}_{n}\rangle\langle r_{n}|$, $L_{R^{2}r}^{(n)} = \sqrt{0.34\gamma_{\mathrm{BBR}}}|R^{2}_{n}\rangle\langle r_{n}|$, $L_{R^{3}r}^{(n)} = \sqrt{0.16\gamma_{\mathrm{BBR}}}|R^{3}_{n}\rangle\langle r_{n}|$, and $L_{R^{4}r}^{(n)} = \sqrt{0.33\gamma_{\mathrm{BBR}}}|R^{4}_{n}\rangle\langle r_{n}|$, with a total BBR rate $\gamma_{\mathrm{BBR}}/(2\pi) = 0.504~\mathrm{kHz}$. Although the Rydberg states populated via BBR transitions from $|r\rangle$ are still subject to spontaneous emission and BBR effects, we consider the probability of such events to be negligible over the duration of the coherent excitation stage in each stabilizer pumping process $S_i^P$.
The population leaking into these states can be transferred to the state $|p\rangle$ via an intermediate Rydberg state $|78S_{1/2}\rangle$ through two resonant transitions: a microwave transition from the leaked Rydberg state to $|78S_{1/2}\rangle$ and an optical transition from $|78S_{1/2}\rangle$ to $|p\rangle$. Subsequently, this BBR-induced leakage is effectively recycled to the ground state $|0\rangle$, analogous to the engineered dissipative reset described above. 

\begin{figure}
	\centering
	\includegraphics[width=0.64\linewidth]{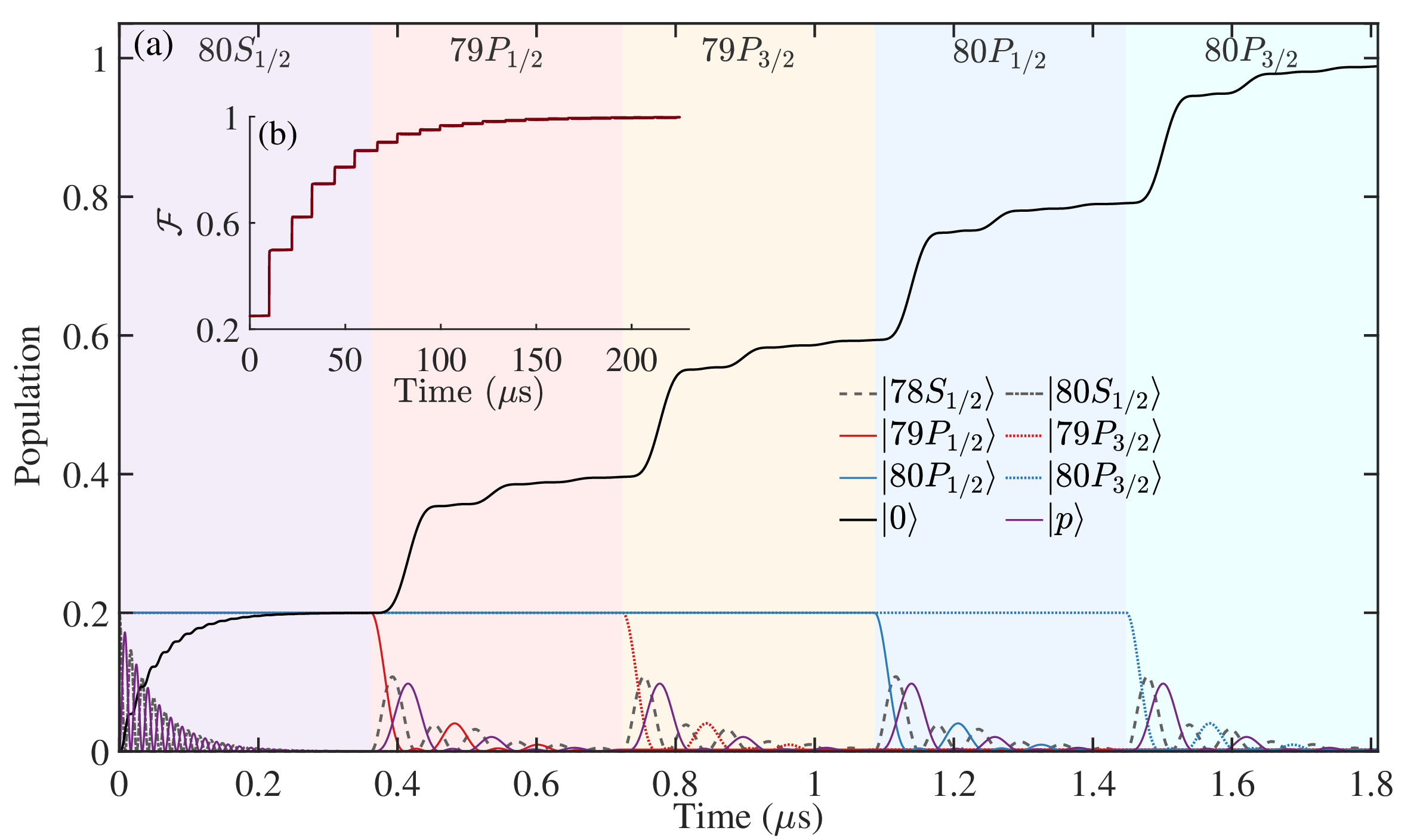}
	\caption{Mitigation of BBR errors at room temperature ($300~\rm K$). 
(a) Demonstration of BBR suppression in a single-atom system. The scheme is designed to sequentially deplete the populations of the Rydberg states $|80S_{1/2}\rangle$, $|79P_{1/2}\rangle$, $|79P_{3/2}\rangle$, $|80P_{1/2}\rangle$, and $|80P_{3/2}\rangle$. 
(b) Evolution of the Bell state fidelity at regular point $\rm K'$. Simulation parameters are $\Omega_a = 2\pi\times 5~\mathrm{MHz}$, $\mathcal{N}=100$, $\alpha = 0.6$, $t_b = \mathcal{N}t_a = 0.1~\mu\text{s}$, $\mathbb{N} = 10$, $\Omega \approx 2\pi\times 13.31~{\rm MHz}$, $\gamma_p = 2\pi\times 6.07~{\rm MHz}$, and decay duration $t_d = 0.362~\mu{\rm s}$. The spontaneous emission and BBR rates from state $|r\rangle$ are $\gamma_{r} = 2\pi\times 0.28~{\rm kHz}$ and $\gamma_{\rm BBR} = 2\pi\times 0.504~{\rm kHz}$, respectively. The coupling strength for the $|r\rangle \leftrightarrow |p\rangle$ transition is set to $10\gamma_p$, while couplings for other transitions are set to $2\gamma_p$ to maintain experimental feasibility within the microwave regime.}\label{fig_BBR}
\end{figure}

To ensure high fidelity, the four BBR-induced leakage states are recycled sequentially in four additional steps controlled by distinct beams, following the main engineered dissipative reset of the Rydberg state \(|r\rangle\). As illustrated in Fig.~\ref{fig_BBR}(a), including the main reset of \(|r\rangle\), the five Rydberg-state populations are sequentially recycled to the ground state \(|0\rangle\) within a total duration of \(1.81~\mu\mathrm{s}\), corresponding to \(t_d=0.362~\mu\mathrm{s}\) per reset step. Specifically, we set the coupling strength between $|r\rangle$ and $|p\rangle$ to $10\gamma_p$, while other couplings are maintained at $2\gamma_p$ to remain within experimental microwave limits. Starting from an initial state $\rho(0) = (|r\rangle\langle r| + \sum_{i=1}^4 |R^i\rangle\langle R^i|)/5$, the final population of the ground state $|0\rangle$ in the single-atom case reaches $98.81\%$, with the potential for even higher population recovery using extended dissipation times. Furthermore, numerical simulations for Bell-state stabilizer pumping [Fig.~\ref{fig_BBR}(b)] yield a fidelity of $99.79\%$ when accounting for spontaneous emission to states $|0\rangle$ and $|1\rangle$.

\section{Experimental proposal}\label{app:experiment}

\begin{figure}
	\centering
	\includegraphics[width=1\linewidth]{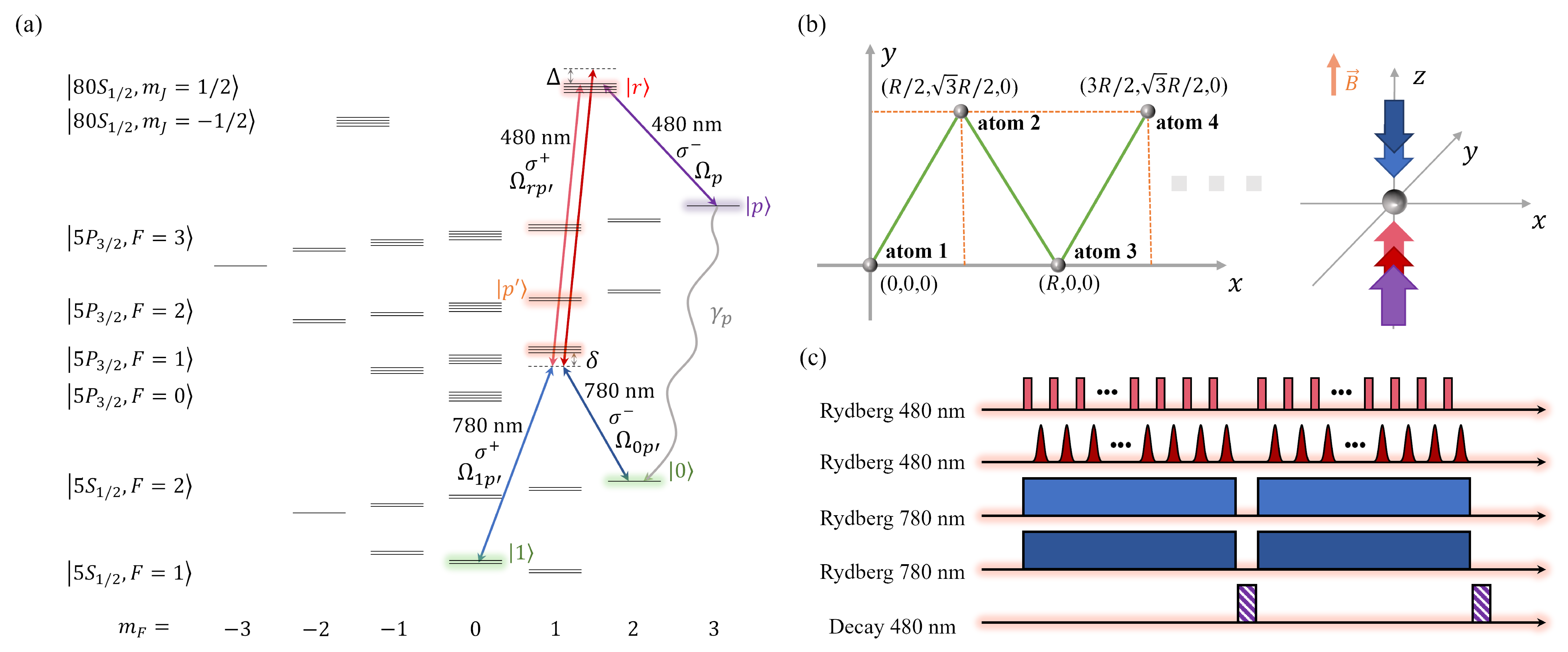}
	\caption{Schematic diagrams illustrating the experimental setup. (a) Coherent excitation and dissipative reset processes for a single atom. (b) Multi-atom arrangement and laser polarization direction for a single atom. (c) Pulse driving sequence, with the two 480 nm laser pulses repeated \(\mathcal{N}\) times, as shown in a representative sketch. The colors in the subfigures consistently represent the distinct characteristics of the pulses.} \label{fig_doppler}
\end{figure}

We provide a detailed description of the experimental implementation. Our approach has two primary objectives: first, to employ purely circularly polarized light to ensure uniform atomic phases and simplify the system dynamics; second, to maximize dissipation toward the \( |0\rangle \) state. The Hamiltonians in Eqs.~(\ref{eq:H_S1})--(\ref{eq:H_Sk}) indicate that the \( |1_i\rangle \) component is volatile, whereas the \( |0_i\rangle \) component remains relatively quiescent. By engineering dissipation directly to the \( |0_i\rangle \) state, we accelerate stabilizer-pumping convergence toward the target stabilizer state.

To this end, we select the stretched ground state \( |0\rangle \equiv |5S_{1/2}, F=2, m_F=2\rangle \), the clock ground state \( |1\rangle \equiv |5S_{1/2}, F=1, m_F=0\rangle \), and the Rydberg state \( |r\rangle \equiv |80S_{1/2}, m_J=1/2\rangle \).
For two-photon excitation, we utilize an intermediate state \( |p'\rangle \equiv |5P_{3/2}, F=1,2,3, m_F=1\rangle \), while for engineered dissipation, we use another intermediate state \( |p\rangle \equiv |5P_{3/2}, F=3, m_F=3\rangle \). We excite the ground state $|0\rangle$ to the Rydberg state $|r\rangle$ via $\sigma^-$- and $\sigma^+$-polarized lights with an intermediate state $|5P_{3/2}, F=1,2,3, m_F=1\rangle$, and excite the ground state $|1\rangle$ to the Rydberg state $|r\rangle$ via two $\sigma^+$-polarized lights with the intermediate state $|5P_{3/2}, F=1,2, m_F=1\rangle$. The component $|m_J=-1/2,m_I=3/2\rangle$ of the states $|5P_{3/2}, F=1,2,3, m_F=1\rangle$ can couple to the Rydberg state $|r\rangle = |80S_{1/2}, m_J=1/2, m_I=3/2\rangle$ due to $\sigma^+$-polarized light and electric dipole interaction. Thus, the Rydberg state $|r\rangle$ will uniquely couple to the stretched state $|p\rangle$ when using $\sigma^-$-polarized light for the engineered dissipative reset.

Specific details, including polarization and wavelength, are depicted in Figs.~\ref{fig_doppler}(a) and \ref{fig_doppler}(b). The pulse sequence is illustrated in Fig.~\ref{fig_doppler}(c), where the driving fields with Rabi frequencies \( \Omega_{0p'} \) and \( \Omega_{1p'} \) are selectively activated according to the stabilizer generator being pumped.

\section{Uniqueness of the stabilizer steady state}\label{sec_steady}

\subsection{Uniqueness of the 1D $N$-qubit Cluster State governed by stabilizer generators $S$}

First, we prove that a stabilizer group can stabilize the cluster states.
The stabilizer generators for a 1D $N$-qubit cluster state are given by the set $\mathcal{S}$
\begin{equation}
    \mathcal{S} = \{ S_1=X_1 Z_2, \quad S_k=Z_{k-1}X_k Z_{k+1} \text{ for } k=2,\dots,N-1, \quad S_N=Z_{N-1}X_N \}.
\end{equation}

The $N$ stabilizer generators $S_i$ are mutually commuting and independent. They generate an Abelian group, known as the stabilizer group $\mathcal{G}$, which contains all possible products of the generators.
\begin{equation}
    \mathcal{G} = \left\{ \prod_{i \in K} S_i \mid K \subseteq \{1, \dots, N\} \right\}.
\end{equation}
This group has an order (number of elements) of $|\mathcal{G}| = 2^N$.

To find the common `+1' eigenspace (the target stabilizer manifold) of all these stabilizers, we can construct a projection operator $\Pi_{\text{stable}}$ by averaging all the elements of the group $\mathcal{G}$
\begin{equation}
    \Pi_{\text{stable}} = \frac{1}{|\mathcal{G}|} \sum_{g \in \mathcal{G}} g = \frac{1}{2^N} \sum_{g \in \mathcal{G}} g.
\end{equation}
This operator projects any arbitrary state in the Hilbert space $\mathcal{H}$ (of dimension $2^N$) onto the stabilizer subspace, which is the subspace of all states $|\psi\rangle$ satisfying $g|\psi\rangle = |\psi\rangle$ for all stabilizers $g \in \mathcal{G}$.

The dimension of the stabilizer subspace is equal to the rank of the projection operator, $\text{rank}(\Pi_{\text{stable}})$. For any projection operator, its rank is equal to the trace. Therefore, we can find the dimension of the subspace by calculating the trace of $\Pi_{\text{stable}}$
\begin{equation}
    \text{rank}(\Pi_{\text{stable}}) = \text{Tr}(\Pi_{\text{stable}}) = \text{Tr}\left( \frac{1}{2^N} \sum_{g \in \mathcal{G}} g \right) = \frac{1}{2^N} \sum_{g \in \mathcal{G}} \text{Tr}(g).
\end{equation}

To evaluate this sum, we consider the trace of each element $g \in \mathcal{G}$. Each element is a Pauli string (a tensor product of Pauli matrices and the identity).

For the identity element $g = \mathcal{I}$, its trace is equal to the dimension of the Hilbert space
\begin{equation}
    \text{Tr}(\mathcal{I}) = 2^N.
\end{equation}

For any other element $g \neq \mathcal{I}$, $g$ is a non-trivial Pauli string. The trace of any non-trivial Pauli matrix is zero, and thus the trace of any non-trivial Pauli string is also zero
\begin{equation}
    \text{Tr}(g) = 0 \quad \text{for } g \neq \mathcal{I}.
\end{equation}

Substituting these results back into the expression for the rank, we find that all terms in the summation vanish except for the identity term
\begin{equation}
    \text{rank}(\Pi_{\text{stable}}) = \frac{1}{2^N} \left[ \text{Tr}(\mathcal{I}) + \sum_{g \in \mathcal{G}, g \neq \mathcal{I}} \text{Tr}(g) \right] = \frac{1}{2^N} (2^N + 0) = 1.
\end{equation}

Since the rank of the projection operator $\Pi_{\text{stable}}$ is 1, the stabilizer subspace is one dimensional. This proves that there exists a unique (up to a global phase) quantum state, the 1D $N$-qubit cluster state $|C_{N}\rangle$, which is simultaneously stabilized by all generators $S_i$.

\subsection{Analysis of the Realistic Stabilizer-Pumping Scheme}

The proof above confirms a fundamental principle: the stabilizer set $\mathcal{S}$ defines a unique quantum state. It provides the theoretical justification for why the $N$-qubit cluster state $|C_N\rangle$ is the target steady state of the stabilizer-pumping dynamics.

However, this elegant group-theoretic proof relies on an idealized assumption: the physical pumping process completely ``attacks" the entire negative eigenspace of every stabilizer generator. However, this may not be the case in physical implementations related to experiments. Defective attack on the negative eigenspace can break the perfect symmetry of the stabilizer group that underpins the projector formalism. Consequently, the system dynamics can no longer be described by the simple group projector $\Pi_{\text{stable}}$. As detailed in our main proposal, our scheme is designed such that the pumping of the bulk stabilizers' negative eigenspaces is unavoidably incomplete. Specifically, for a stabilizer like $S_k = Z_{k-1}X_k Z_{k+1}$, the state $|1_{k-1} -_k 1_{k+1}\rangle$ is not directly excited. The previous proof, while correct for the ideal case, is insufficient to guarantee the success of our realistic scheme. Consequently, we present a more robust and generalized proof, which provides a fundamental theoretical explanation for the practical extension of further pumping and dissipation type schemes.

To rigorously analyze this more realistic scenario, we adopt a more fundamental and robust framework---system's kernel space, to directly model the physical stabilizer-pumping scheme. This approach is more powerful because it relies only on the specific set of states $\{|j\rangle\}$ that are pumped by the lasers.

Define a total ``pump operator'' $\mathcal{P}$, which encodes the subspace directly addressed by the coherent pumping fields. It is constructed from the projectors onto all states $|j\rangle$ that are excited
\begin{equation}
    \mathcal{P} = \sum_{j} P_j = \sum_{j} |j\rangle\langle j|.
\end{equation}
A quantum state $|\psi\rangle$ is a dark state of the pumping step if and only if it is completely unaffected by the pumping. This means that it must lie in the null space of the pump operator, denoted $\ker(\mathcal{P})$:
\begin{equation}
    \ker(\mathcal{P}) = \{ |\psi\rangle \in \mathcal{H} \mid \mathcal{P} |\psi\rangle = \mathbf{0} \}.
\end{equation}
Since all $P_j$ are positive semi-definite, this condition is equivalent to the requirement that $\langle j | \psi \rangle = 0$ for all pumped states $|j\rangle$.

Therefore, the existence of a unique dark state of the pumping step is guaranteed if and only if the dimension of this kernel is one:
\begin{equation}
    \dim[\ker(\mathcal{P})] = 1.
\end{equation}

We now apply this formalism to practical schemes. The central issue can be expressed as follows: even with this reduced set of directly pumped states, is the dimension of the resulting kernel still one?

By applying the Rank-Nullity Theorem, the dimension of the kernel can be given by
\begin{equation}
    \dim[\ker(\mathcal{P})] = \dim(\mathcal{H}) - \text{rank}(\mathcal{P}).
\end{equation}
The rank of $\mathcal{P}$ is the number of linearly independent vectors in the set of pumped states $\{|j\rangle\}$.

As encountered in our scheme, even with the exclusion of the $|1-1\rangle$-type states, the set of pumped states is sufficiently rich to ensure that it contains $2^N - 1$ linearly independent vectors. Consequently, the rank of the pump operator is $\text{rank}(\mathcal{P}) = 2^N - 1$. Therefore,
\begin{equation}
    \dim[\ker(\mathcal{P})] = 2^N - (2^N - 1) = 1.
\end{equation}
This proves that despite the incomplete direct pumping, our scheme still possesses a unique dark state of the pumping dynamics. Finally, since we know by construction that the target cluster state $|C_N\rangle$ is orthogonal to all pumped states (as it is a +1 eigenstate of all stabilizers), it must be an element of the kernel. Given that the kernel is one-dimensional, it uniquely identifies $|C_N\rangle$ as the dark state of the pumping dynamics. We can thus state our final conclusion with mathematical precision as follows:
\begin{equation}
    \ker(\mathcal{P}) = \text{span}\{|C_N\rangle\}.
\end{equation}


\subsection{The Pump Operator as a Constraint System: Uniqueness, Robustness, and Speed}

Stabilizer pumping toward a target stabilizer state can be viewed as the application of a set of constraints on the Hilbert space of the system. The pump operator, $\mathcal{P} = \sum_{j} \ket{j}\bra{j}$, defines this set, and a state $\ket{\psi}$ survives as a steady dark state only if it is orthogonal to all pumped states, i.e., $\braket{j|\psi}=0$ for all $\ket{j}$. The success of a stabilizer-pumping scheme therefore depends critically on whether the implemented constraints are sufficient to uniquely define the target stabilizer state, and the nature of these constraints dictates the pumping convergence rate.

(i) The Necessity of a Sufficient Constraint Set. 
An insufficient set of constraints fails to produce a unique steady state. If the subspace spanned by the pumped states is too small, the kernel of the pump operator will have a dimension greater than one, allowing for multiple, unwanted dark states. A clear example of this failure is found in a naive stabilizer-pumping construction for a two-qubit cluster state, stabilized by $S_1=X_1Z_2$ and $S_2=Z_1X_2$. If one implements an incomplete pumping scheme that only imposes two constraints---for instance, by only pumping the states $\ket{-0}$ and $\ket{0-}$---the dark state subspace is two-dimensional: one is $|C_2\rangle$ and another is $1/\sqrt{3}(|C_{2}^{+-}\rangle + |C_{2}^{-+}\rangle + |C_{2}^{--}\rangle)$ composed of three other orthogonal states. This scheme fails because the constraints are inadequate to eliminate all unwanted dark-state components orthogonal to the target two-qubit cluster state.

(ii) Robustness via Indirect and Overlapping Constraints.
Our realistic scheme for stabilizer pumping toward the cluster state avoids this failure, even though it employs an incomplete set of direct constraints. The key to its success lies in the power of indirect, overlapping constraints. A ``constraint" is fundamentally a mathematical condition of non-orthogonality: any pumped state $\ket{j}$ constrains any state $\ket{\psi}$ as long as their inner product $\braket{j|\psi}$ is non-zero. For instance, in the three-qubit scheme, we do not directly pump the state $\ket{1_1-_21_3}$. However, this state is still constrained by the implemented pumping, such as the projectors $\ket{-_{1}0_{2}}\bra{-_{1}0_{2}}\otimes \mathcal{I}_{3}$ and $ \mathcal{I}_{1} \otimes \ket{0_{2}-_{3}}\bra{0_{2}-_{3}}$, which have non-zero overlaps with $\ket{1_1-_21_3}$. This non-zero inner product constitutes a valid, albeit indirect, constraint that prevents $\ket{1_1-_21_3}$ from being a dark state. The robustness of our scheme stems from the fact that the full set of these indirect constraints is sufficient to compensate for the missing direct ones, ensuring a unique steady state.

(iii) Reduced Direct Pumping Channels and Convergence Rate.
Finally, while these overlapping constraints guarantee uniqueness, the reduction in the total number of direct pumping pathways adversely affects the stabilizer-pumping convergence rate. This performance degradation can be explained by analyzing the system gap of the Floquet Liouvillian $\mathcal{L}_F$, given by $g_{F} = \min_{\lambda'_k \neq 0} |\text{Re}[\lambda'_k]|$, where $\lambda'_k$ are the eigenvalues of $\mathcal{L}_F$. Each directly pumped state $\ket{j}$ provides an effective dissipative pathway in the Floquet-Liouvillian dynamics. In an ideal case where all negative eigenspaces of all stabilizers are pumped, the Floquet Liouvillian $\mathcal{L}_{F_{\text{ideal}}}$ possesses a certain gap $g_{F_{\text{ideal}}}$. In our realistic scheme, the failure to excite some states (such as $\ket{1_{k-1}-_k 1_{k+1}}$) means that our Floquet Liouvillian $\mathcal{L}_{F_{\text{real}}}$ is missing these corresponding direct pumping pathways. The removal of these direct decay pathways can be shown to reduce the overall magnitude of dissipation, leading to a smaller dissipative gap, i.e., $g_{F_{\text{real}}} \le g_{F_{\text{ideal}}}$. Intuitively, population in stabilizer-violating sectors that could have rapidly decayed through a missing direct channel must now follow more complex, multi-step paths to be dissipated by the remaining channels. These slower, indirect pathways are associated with eigenvalues closer to zero, thus shrinking the gap. 

\begin{figure}
	\centering
	\includegraphics[width=18cm]{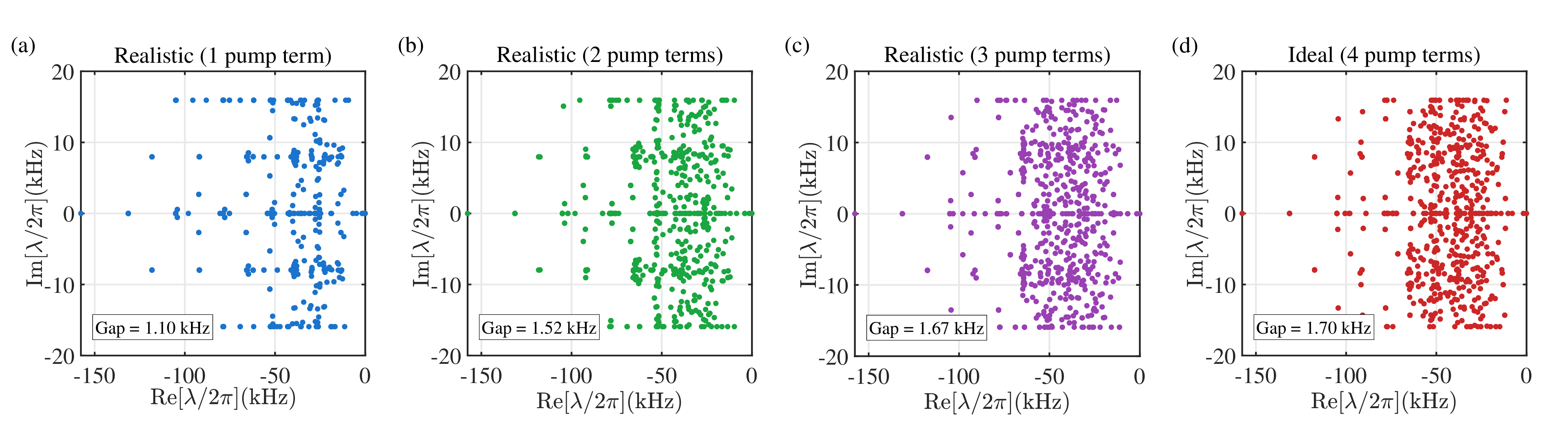}
	\caption{Floquet-Liouvillian spectrum for stabilizer pumping toward the three-qubit cluster state. According to the pumping of stabilizer $Z_{1}X_{2}Z_{3}$, it is divided into four situations: (a) one pump state ($|0-0\rangle$), (b) two pump states ($|1+0\rangle, |0-0\rangle$), (c) three pump states ($|0+1\rangle, |1+0\rangle, |0-0\rangle$), and (d) four pump states ($|0+1\rangle, |1+0\rangle, |0-0\rangle, |1-1\rangle$). These one, two, and three pump terms correspond to realistic incomplete pumping, while four pump terms correspond to ideal complete pumping. Here the coherent Floquet-excitation stage is modeled using the effective Hamiltonian derived in Appendix~\ref{sec:eff_Ham}. Parameters are $\Omega_a = 2\pi\times 5~\mathrm{MHz}$, $\mathcal{N}=100$, $\alpha = 0.6$, and $t_b = \mathcal{N}t_a = 0.1~\mu\text{s}$, $\Omega \approx 2\pi\times 13.31~{\rm MHz}$, and decay duration $t_d = 0.362~\mu{\rm s}$.}\label{Cluster3_LF_QES}
\end{figure}

A prime example is the stabilizer pumping of a three-qubit cluster state. According to stabilizer $Z_{1}X_{2}Z_{3}$, we consider four cases of stabilizer pumping, including one pump state ($|0-0\rangle$), two pump states ($|1+0\rangle, |0-0\rangle$), three pump states ($|0+1\rangle, |1+0\rangle, |0-0\rangle$), and four pump states ($|0+1\rangle, |1+0\rangle, |0-0\rangle, |1-1\rangle$). The Floquet-Liouvillian spectra are shown in Fig.~\ref{Cluster3_LF_QES}, which all have the same and unique zero-eigenvalue steady-state mode $|r_{0}\rangle\rangle = |C_{3}\rangle \otimes |C_{3}\rangle$ satisfying $\mathcal{L_{F}}|r_{0}\rangle\rangle = 0$. These cases (one-, two-, and three-pump), lacking a direct decay channel via the $S_2$ pump, are instead targeted by the pumps for the adjacent stabilizers $S_1=X_1Z_2$ and $S_3=Z_2X_3$. This forces its population into a slower, multi-step decay process, ultimately shrinking the dissipative gap. Consequently, the stabilizer-pumping convergence time, which scales as $T_{\text{prep}} \sim 1/g_{F}$, is increased. Figure~\ref{Cluster3_LF_QES} displays that the gap decreases with the reduction of pump terms. This analysis confirms that while our scheme is robust in reaching the target stabilizer state, it comes at the cost of reduced stabilizer-pumping efficiency.

\section{Parallel stabilizer pumping for accelerated cluster-state stabilization}\label{app:parallel}

Reference~\cite{Bluvstein2022A} reports the experimental preparation of multipartite 1D cluster states through a carefully crafted CZ gate scheme. 
By leveraging the mobility of atoms manipulated via optical tweezers, this approach enables cluster state generation in the minimal number of steps, albeit with a modest reduction in fidelity. This method can be seamlessly integrated into our framework as a rapid initialization step prior to stabilizer pumping. Nevertheless, we propose an alternative approach with lower cost to accelerate the convergence toward multipartite cluster states through parallel stabilizer pumping. 

\begin{figure}
	\centering
	\includegraphics[width=11.5cm]{fig_11pumping_order}
	\caption{Schematic of the parallel stabilizer-pumping sequence for an eight-qubit cluster state, executed as follows: (i)~$S_2$ and $S_7$, (ii)~$S_3$ and $S_8$, (iii)~$S_4$, (iv)~$S_5$, (v)~$S_1$ and $S_6$.}\label{fig_pumping_order}
\end{figure}

\begin{figure}
	\centering
	\includegraphics[width=18cm]{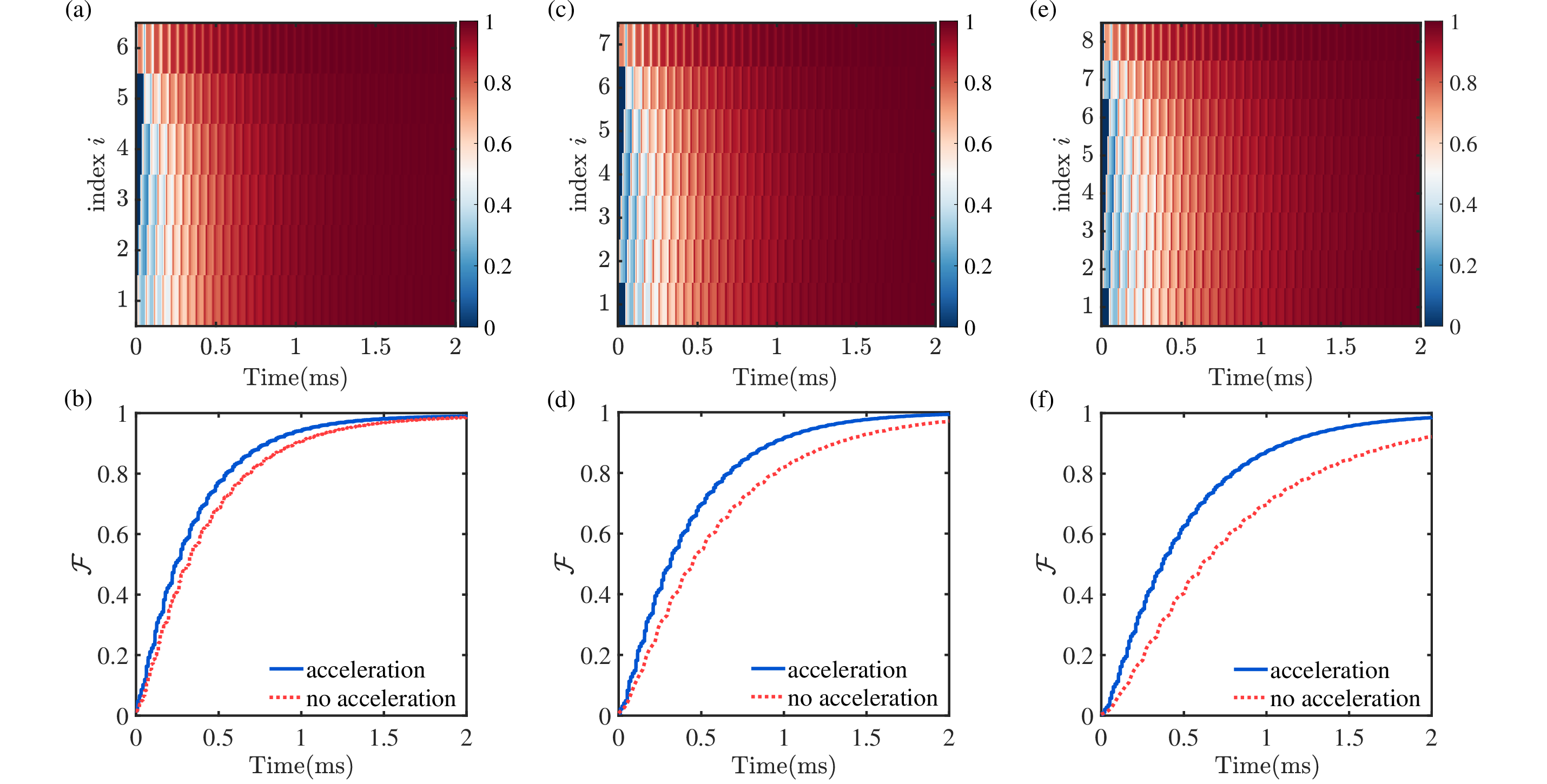}
	\caption{Accelerated stabilizer pumping toward six- (a,b), seven- (c,d), and eight-qubit cluster states (e,f). (a), (c) and (e) display the temporal evolution of the expectation value of each stabilizer generator $\langle S_i\rangle$. (b), (d) and (f) display the comparison of fidelity with and without parallel stabilizer pumping. Parameters are $\Omega_a/(2\pi) = 5~\mathrm{MHz}$, $\Delta=U_{rr}=77\Omega_a$, $\mathcal{N}=100$, $\alpha = 0.6$, $t_b = \mathcal{N}t_a = 0.1~\mu{\rm s}$, $\Omega \approx 2\pi\times 13.31~{\rm MHz}$, decay duration $t_d = 0.362~\mu{\rm s}$, and distance $R = 4.7171~\mu{\rm m}$.}\label{accrelate_Cluster_678}
\end{figure}

Our proposed scheme divides the three-qubit bulk stabilizer-pumping operations into five sequential steps: (i)~$S_{2+5(n-1)}$, (ii)~$S_{3+5(n-1)}$, (iii)~$S_{4+5(n-1)}$, (iv)~$S_{5+5(n-1)}$, and (v)~$S_{6+5(n-1)}$, where $n = 1, 2, 3, \dots$. For the two-qubit boundary stabilizer-pumping operations, $S_1$ is incorporated into step (v), while $S_N$ is embedded in the step where the $S_{N-5}$ is located. An example is presented in Fig.~\ref{fig_pumping_order}, which displays the stabilizer-pumping order for the eight-qubit cluster state. To evaluate the acceleration performance, we conducted numerical simulations of stabilizer pumping toward six-, seven-, and eight-qubit cluster states, as shown in Fig.~\ref{accrelate_Cluster_678}. The six-qubit cluster state simulation employs the original Hamiltonian, while the seven- and eight-qubit cluster state simulations utilize the effective Hamiltonian. As the number of atoms increases, the acceleration effect becomes increasingly pronounced.

\section{Extensions of the stabilizer-pumping scheme}\label{app:extensions}

\subsection{Complete stabilizer pumping in one dimension}

First, we note that complete stabilizer-pumping dynamics can also be achieved. For the bulk stabilizers $S_{k} = Z_{k-1}X_{k}Z_{k+1}$, we can use a two-step coherent excitation followed by dissipative reset to realize complete stabilizer pumping.

(i) Illuminate atoms with lasers to achieve an effective Hamiltonian described as 
\begin{equation}
    H_{S_{k}}^{(i)} = \frac{\Omega_\text{eff}}{2}(|r_{k-1}+_{k}0_{k+1}\rangle\langle 1_{k-1}+_{k}0_{k+1}| + |0_{k-1}r_{k}0_{k+1}\rangle\langle 0_{k-1}-_{k}0_{k+1}| + |0_{k-1}+_{k}r_{k+1}\rangle\langle 0_{k-1}+_{k}1_{k+1}|) + {\rm H.c.},
\end{equation}
with pulse duration $T_{(i)} = \pi/\Omega_{\text{eff}}$.

(ii) Illuminate atoms with lasers to achieve an effective Hamiltonian described as 
\begin{equation}
    H_{S_{k}}^{(ii)} = \frac{\Omega_\text{eff}}{2}(|1_{k-1}r_{k}1_{k+1}\rangle\langle 1_{k-1}-_{k}1_{k+1}|) + {\rm H.c.},
\end{equation}
with the same pulse duration $T_{(ii)} = \pi/\Omega_{\text{eff}}$. After completing these two steps, our operation sequence realizes complete pumping of the bulk stabilizer $S_{k}$.

The remaining task is to provide an implementation method for an effective Hamiltonian $H_{S_{k}}^{(ii)}$. We can execute these two Hamiltonians 
\begin{eqnarray}
    H_{a_{(ii)}}^{S_{k}}(t) &=& \frac{\Omega_{a}}{2}(\mathcal{I}_{k-1}\otimes|r_{k}\rangle\langle -_{k}|\otimes\mathcal{I}_{k+1}) + {\rm H.c.} + H_{V},\cr\cr
    H_{b_{(ii)}}^{S_{k}}(t) &=& \frac{\Omega_{b}(t)}{2}e^{-i\Delta t}(|r_{k-1}\rangle\langle 0_{k-1}|\otimes\mathcal{I}_{k}\otimes\mathcal{I}_{k+1} + \mathcal{I}_{k-1}\otimes\mathcal{I}_{k}\otimes|r_{k+1}\rangle\langle 0_{k+1}|) + {\rm H.c.} + H_{V}.\nonumber
\end{eqnarray}
At this point, we can achieve complete stabilizer pumping in the idealized construction.

\subsection{Stabilizer pumping toward $N$-qubit GHZ states}

\begin{figure}
	\centering
	\includegraphics[width=7cm]{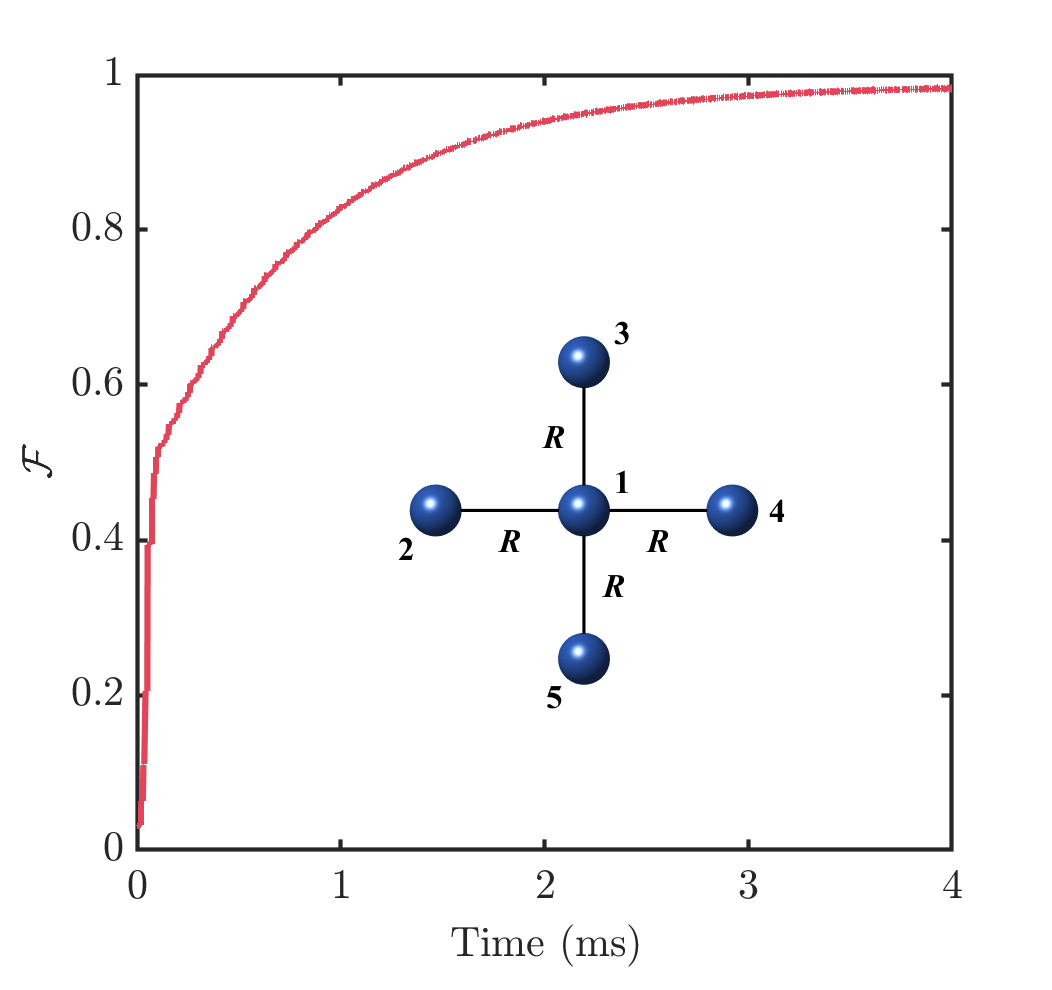}
	\caption{Fidelity of the five-qubit GHZ (star-like graph) state during stabilizer pumping. The inset shows the geometric structure of the five-qubit GHZ state. Parameters are $\Omega_a = 2\pi\times 5~\mathrm{MHz}$, $\Delta=U_{rr}=77\Omega_a$, $\mathcal{N}=100$, $\alpha = 0.6$, $t_b = \mathcal{N}t_a = 0.1~\mu\text{s}$, $\Omega \approx 2\pi\times 13.31~{\rm MHz}$, decay duration $t_d = 0.362~\mu{\rm s}$, and nearest-neighboring distance $R = 4.7171~\mu{\rm m}$.}\label{GHZ5_original}
\end{figure}

This framework can be naturally generalized to stabilizer pumping toward $N$-qubit entangled GHZ states. Here we provide the stabilizer definition
\begin{equation} \label{eq:ghz_stabilizers}
    S_1^{(\text{GHZ}_N)} = \prod_{i=1}^{N} X_{i}, \quad  S_k^{(\text{GHZ}_N)} = Z_{1}Z_{k}, \quad \text{for } k = 2, 3, \dots, N.
\end{equation}
This requires the first atom to be at the center of a circle, while the other atoms are on the edges of the circle (star-shaped structure). Ideally, the distances between neighboring atoms on the circle should be equal. As an example, we display the stabilizer-pumping implementation for the five-qubit GHZ state as shown in the illustration of Fig.~\ref{GHZ5_original}
\begin{eqnarray}\label{eq5}
     |\mathcal{G}_{5}\rangle = \frac{1}{\sqrt{2}}(|00000\rangle + |11111\rangle).
\end{eqnarray}
It can be stabilized by generators $\{X_{1}X_{2}X_{3}X_{4}X_{5}, Z_{1}Z_{2}, Z_{1}Z_{3}, Z_{1}Z_{4}, Z_{1}Z_{5}\}$ with eigenvalue $+1$ for each generator.
We mainly display an abbreviated scheme for the five-body stabilizer $X_{1}X_{2}X_{3}X_{4}X_{5}$, which can be realized by an effective Hamiltonian
\begin{equation}
    H_{\mathcal{G}_5} = \frac{\Omega_{\rm eff}}{2}|r++++\rangle\langle -++++| + {\rm H.c.}
\end{equation}
Similar to the discussion in the preceding subsection, the effective Hamiltonian can be achieved by
\begin{eqnarray}
    H_{a}^{{\mathcal{G}_5}}(t) &=& \frac{\Omega_{a}}{2}(|r_{1}\rangle\langle -_{1}| \otimes \mathcal{I}_{2} \otimes\mathcal{I}_{3} \otimes\mathcal{I}_{4} \otimes\mathcal{I}_{5}) + {\rm H.c.} + H_{V},\cr\cr
    H_{b}^{{\mathcal{G}_5}}(t) &=& \frac{\Omega_{b}(t)}{2}e^{-i\Delta t}(\mathcal{I}_{1} \otimes |r_{2}\rangle\langle -_{2}| \otimes \mathcal{I}_{3}\otimes\mathcal{I}_{4}\otimes\mathcal{I}_{5}
    + \mathcal{I}_{1}\otimes\mathcal{I}_{2}\otimes|r_{3}\rangle\langle -_{3}|\otimes\mathcal{I}_{4}\otimes\mathcal{I}_{5}\cr\cr
    &&+ \mathcal{I}_{1}\otimes\mathcal{I}_{2}\otimes\mathcal{I}_{3}\otimes|r_{4}\rangle\langle -_{4}|\otimes\mathcal{I}_{5} + \mathcal{I}_{1}\otimes\mathcal{I}_{2}\otimes\mathcal{I}_{3}\otimes\mathcal{I}_{4}\otimes|r_{5}\rangle\langle -_{5}|) + {\rm H.c.} + H_{V}.\nonumber
\end{eqnarray}

The first atom is resonantly excited to the Rydberg state $|r\rangle$ first and then drives the other four atoms far off-resonantly. Although this excitation scheme can only target a single negative eigenstate of the five-body stabilizer $X_1X_2X_3X_4X_5$, it remains a viable approach. However, the trade-off is that the stabilizer-pumping convergence toward the GHZ state becomes slower. Figure~\ref{GHZ5_original} depicts the fidelity of the five-qubit GHZ state.

\subsection{Stabilizer pumping toward two-dimensional cluster states}

\begin{figure}
 	\centering
 	\includegraphics[width=19cm]{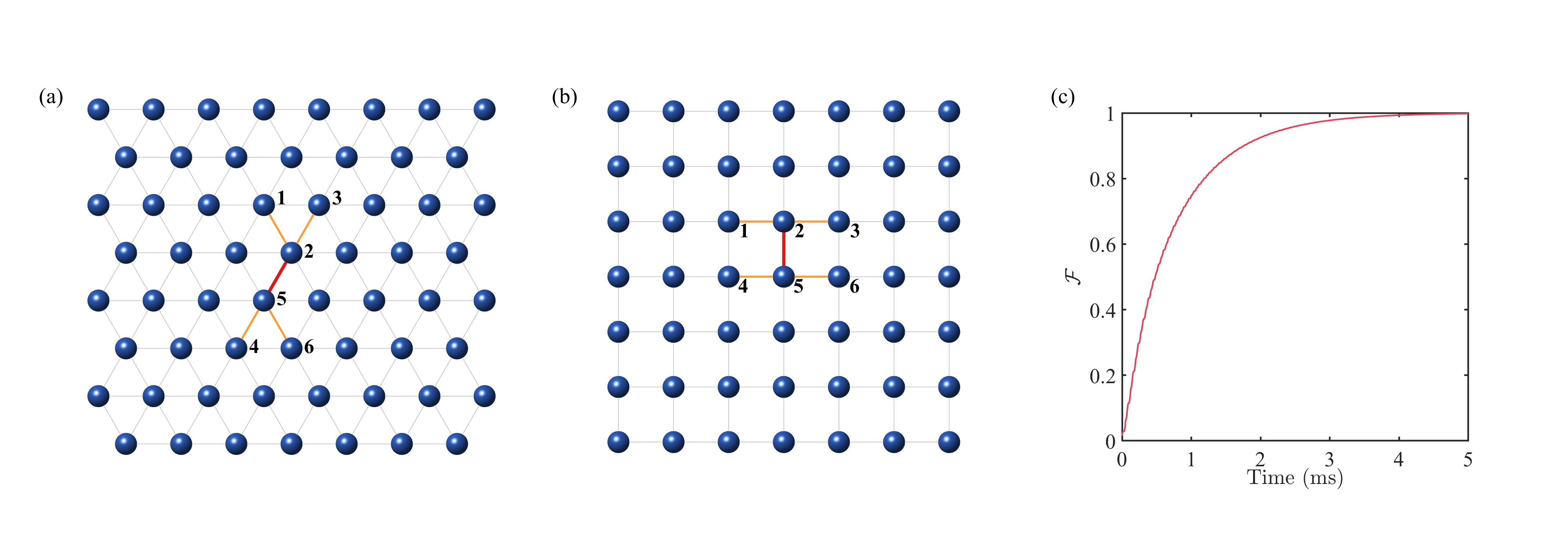}
 	\caption{Structures of 2D cluster states formed by triangular lattice (a) and square lattice (b). The 2D six-qubit cluster states are marked by a yellow line, with the edges connecting the two 1D three-qubit cluster states highlighted in red. (c) Stabilizer pumping toward the 2D six-qubit cluster state using the effective Hamiltonians structured in (a) and (b). Under the consideration of effective excitation, both structures (a) and (b) can achieve perfect excitation of two-qubit stabilizers, while the same effective excitation Hamiltonians in Eq.~(\ref{eq:4_qubit_stabilizer}) are required for four-qubit stabilizers, thus they have the same effective evolution results. The parameters in effective scheme are set as $\Omega_a = 2\pi\times 5~\mathrm{MHz}$, $\mathcal{N}=100$, $\alpha = 0.6$, $t_b = \mathcal{N}t_a = 0.1~\mu\text{s}$, and decay duration $t_d = 0.362~\mu{\rm s}$.}\label{fig_2D_lattice}
 \end{figure}

In the main text, we have considered stabilizer pumping toward a 1D cluster state and a graph-state resource for a CNOT operation, which meet the basic requirements of one-way quantum computing. Building on these results, we now extend our framework to stabilizer pumping toward two-dimensional (2D) cluster states by integrating the aforementioned theoretical protocols. As shown in Figs.~\ref{fig_2D_lattice}(a) and~\ref{fig_2D_lattice}(b), we separately propose 2D cluster states with structures of triangular lattice and square lattice. Using a six-qubit cluster state with stabilizer generators $\{X_{1}Z_{2}, Z_{1}X_{2}Z_{3}Z_{5}, Z_{2}X_{3}, X_{4}Z_{5}, Z_{2}Z_{4}X_{5}Z_{6}, Z_{5}X_{6}\}$ as a representative example, we demonstrate the stabilizer-pumping strategy for both triangular and square lattice geometries.

In fact, the 2D six-qubit cluster state only requires two- and four-qubit stabilizer pumping, meaning that both structures require the same operations. A combination of at least three excitation terms is required for the two four-qubit stabilizers to ensure a unique dark state (i.e., maintaining sufficient constraints). One possible choice is
\begin{eqnarray}
    H^0_{|-_2\rangle\leftrightarrow|r_2\rangle} = \frac{\Omega_{\rm eff}}{2}|0_{1}r_{2}0_{3}0_{5}\rangle\langle 0_{1}-_{2}0_{3}0_{5}|+\text{H.c.},\cr\cr
   H^1_{|+_2\rangle\leftrightarrow|r_2\rangle} = \frac{\Omega_{\rm eff}}{2}|1_{1}r_{2}1_{3}1_{5}\rangle\langle 1_{1}+_{2}1_{3}1_{5}|+\text{H.c.},\cr\cr
    H^0_{|-_5\rangle\leftrightarrow|r_5\rangle}= \frac{\Omega_{\rm eff}}{2}|0_{2}0_{4}r_{5}0_{6}\rangle\langle 0_{2}0_{4}-_{5}0_{6}|+\text{H.c.}
\label{eq:4_qubit_stabilizer}
\end{eqnarray}
As shown in Fig.~\ref{fig_2D_lattice}(c), the numerical simulation using the effective Hamiltonians demonstrates the feasibility of the scheme.

\twocolumngrid
\bibliography{PRA.bbl}

\end{document}